\documentclass[aip,jcp,preprint]{revtex4-1}
\usepackage{amsmath}
\usepackage{amsfonts}
\usepackage{amssymb}
\usepackage{graphicx}%
\usepackage{color}
\usepackage{xcolor}
\usepackage{subfigure}

\allowdisplaybreaks

\newcommand{\bfm}[1]{\mbox{\boldmath $#1$}}
\graphicspath{{figs/}}

\begin{document}

\title{Large deviations of Rouse polymer chain: first passage problem}
\author{Jing Cao, Jian Zhu, Zuowei Wang, Alexei E. Likhtman}
\affiliation{Department of Mathematics and Statistics, University of Reading, Reading RG6 6AX,UK}

\date{August 2015}%
\revised{xxx 20xx}%

\begin{abstract}

The purpose of this paper is to investigate several analytical methods of solving first passage (FP) problem for the Rouse model, a simplest model of a polymer chain. We show that this problem has to be treated as a multi-dimensional Kramers' problem, which presents rich and unexpected behavior. 
We first perform direct and forward-flux sampling (FFS) simulations,
and measure the mean first-passage time $\tau(z)$ for the free end to reach a certain distance $z$ away from the origin.
The results show that the mean FP time is getting faster if the Rouse chain is represented by more beads. 
Two scaling regimes of $\tau(z)$ are observed, with transition between them varying as a function of chain length. 
We use these simulations results to test two theoretical approaches. One is a well known asymptotic theory valid in the limit of zero temperature. We show that this limit corresponds to fully extended chain when each chain segment is stretched, which is not particularly realistic. A new theory based on the well known Freidlin-Wentzell theory is proposed, where dynamics is projected onto the minimal action path. The new theory predicts both scaling regimes correctly, but fails to get the correct numerical prefactor in the first regime. Combining our theory with the FFS simulations lead us to a simple analytical expression valid for all extensions and chain lengths.
One of the applications of polymer FP problem occurs in the context of branched polymer rheology.
In this paper, we consider the arm-retraction mechanism in the tube model,
which maps exactly on the model we have solved. The results are compared to the Milner-McLeish theory without constraint release,
which is found to overestimate FP time by a factor of 10 or more.

\end{abstract}

\maketitle

\section{Introduction}

Rouse model is the simplest stochastic model of polymer dynamics, where almost
everything can be solved analytically. The polymer chain is modelled with a
set of beads with coordinates $\bfm{R}_{i}$ and friction $\xi_{0}$ experiencing
Brownian motion and connected into the chain by a set of harmonic springs with
spring constant $k=\dfrac{3k_{B}T}{b^{2}}$ where $k_{B}$ and $T$ are Boltzmann
constant and temperature and $b$ is the statistical segment length, respectively. 
Thus, the equation of motion of a given bead $i$ is%
\begin{equation}
\xi_{0}\frac{d\bfm{R}_{i}}{dt}=\frac{3k_{B}T}{b^{2}}\left(  \bfm{R}_{i+1}+\bfm{R}_{i-1}%
-2\bfm{R}_{i}\right)  +f_{i}(t);\quad\left\langle f_{i}(t)\right\rangle
=0;\quad\left\langle f_{i}(t)f_{j}(t^{\prime})\right\rangle =2k_{B}T\xi_{0}\delta_{ij}%
\delta(t-t^{\prime}) \label{eqrouse}%
\end{equation}
where the first term on the right hand side represents the connectivity forces 
and the second term is random white noise force. This equation is valid for all beads $i$
apart from the two end beads. The equation for the end beads depends on the
problem - they can be fixed, free, or have a constant force acting on them. In
this paper we will consider chains with one end fixed at the origin
($\bfm{R}_{0}=\bfm{0})$ and the second end --- free.

Most experimental quantities of interest, such as stress or end-to-end
relaxation functions, can be calculated analytically for Rouse model. This is
done by transforming the system of coupled eq.\ref{eqrouse} into a set of
independent equations for the eigenmodes of this system $\{X_{p}\}$, called
Rouse modes in this context. Equation for each mode can then be solved
independently, and all physical quantities are expressed as combination
of these Rouse modes and their correlation functions.

A notable exception from this is the first-passage (FP) problem (often called
Kramers' problem), which is still not solved analytically even for the Rouse
model. The FP problem requires calculation of the average time that the free end of polymer
chain reaches a particular point for the first time after starting from some
specified initial set of conformations (e.g. those in equilibrium state). 
This problem appears
in many areas of polymer physics, ranging from biophysics\cite{Marques:2001} to
chemical reactions\cite{deGennes:1982,Leibler:1996,Fredrickson:1997} to polymer rheology\cite{Milner:1997,Milner:1998}.
It is also related to protein folding, although in proteins all amino acids must
fold into correct places, not just the end groups. 
In this work, we shall
concentrate on a particular example from rheology of polymer stars, but we
believe many of our findings will be applicable to other scientific areas as well
as to general FP problems in multi-dimensional systems.

In polymer rheology, the dynamics of polymer melts made from long chains can
not be described by the Rouse model because the chains get entangled with each other,
which means that their interactions are much more complex than those reflected by
eq.\ref{eqrouse}. This, however, can be approximately solved using the tube model
concept: one assumes that each chain is moving parallel to the contour of the
tube (made by other chains) and the motion perpendicular to the tube is
suppressed. In order to relax imposed deformation, the chain must wiggle out of
the tube into the newly created tube which does not carry the memory about
deformation. If we accept this concept, the problem of stress relaxation in
entangled polymer melt is reduced to the first-passage problem because the stress
associated with each tube segment gets relaxed when the chain end reaches it
for the first time\cite{Doi1988}. Since the motion of the chain along the tube is assumed to
be unaffected by entanglements, one usually assumes that the Rouse model
eq.\ref{eqrouse} is valid if we take $\bfm{R}_{i}$ to be the positions of monomers along
the tube axis. In other words, in order to solve most problems in entangled
polymer dynamics, one has to start by solving FP problem for the
one-dimensional Rouse chain, which is exactly the focus of this paper. We note
that entangled polymer dynamics is greatly complicated by the fact that the
tube itself is not fixed in space due to the motion of other chains surrounding the
target chain, a phenomenon which is called constraint release. 
We will not address this problem in the current paper, or in other
words we will assume that surrounding chains are much longer than our target
chain and therefore provide permanent network of entanglements.

Before plunging into the depth of the problem, we briefly describe the
Milner-McLeish theory\cite{Milner:1997,Milner:1998} currently used in polymer rheology of stars. 
The theory
assumed that the Rouse chain inside the tube can be replaced by one bead
attached to the origin through a harmonic spring. 
The spring constant was then chosen to be
$k=\frac{3k_{B}T}{Nb^{2}}$such that the average squared spring length is equal
to the mean-square end-to-end distance of the polymer chain (where $N$ is the number of
moving beads in the chain). 
The bead friction was chosen to be $\xi_{eff}=\frac{N}{2}\xi_{0}$, i.e. to carry half of the friction of the whole
chain. Authors argued that the reason for this $1/2$ factor is that the
average bead only needs to travel distance $L/2$ if the end is to travel
distance $L$. 
The FP problem of the one bead model has an exact solution (Kramers formula)%
\begin{equation}
\tau(z)=\frac{\xi_{eff}}{k_B T}\int_0^z dx \exp\Big(\frac{U(x)}{k_B T}\Big)\int_{-\infty}^{x} \exp\Big(-\frac{U(x')}{k_B T}\Big) dx' 
\approx \frac{\xi_{eff}}{U^{\prime}(z)}\sqrt{\frac{2k_B T\pi
}{U^{\prime\prime}(0)}}\exp\left(  \frac{U(z)}{k_B T}\right) \label{eqn:Kramers}
\end{equation}
where $U(z)=\dfrac{3k_B T z^2}{2N b^2} $.
The approximation is valid if $U(z)\gg k_{B}T$.
This leads to the following expression for the mean FP time%
\begin{align}
\tau_{MM}(z)=\frac{\pi^{5/2}}{4\sqrt{6}} \tau_R \frac{\sqrt{N}b}{z} \exp\Big(\frac{3 z^2}{2 N b^2}\Big) \label{eqn:MMTheory}
\end{align}
where $\tau_R=\dfrac{4\xi_0 N^2 b^2}{3\pi^2 k_B T}$ is the longest relaxation time of one-end fixed continuous Rouse chain,
which is $4$ times larger than that of the chain with two free ends.

In this paper we shall first perform direct and forward-flux sampling simulations of
the FP time of one-dimensional Rouse chain and compare them with the Milner-McLeish
assumption (section 2). 
A very significant disagreement will motivate us to
look at the exact asymptotic solution of a multi-dimensional Kramers problem in
section 3, which we will find methodologically useful but invalid for the
realistic extensions of Rouse chain. In section 4 we will develop a new FP
theory for the multi-dimensional Gaussian process, which will result in
correct scaling behaviour but an incorrect prefactor in the intermediate regime. 
We will finally combine
simulation and analytical results into a simple formula for the FP
time of discrete and continuous Rouse chains for any large extension where
$U(z)\gg k_{B}T$.

\section{Direct and Forward Flux sampling simulations}

In order to set the scene for further calculations, and to evaluate the Milner-McLeish ansatz, eq.\ref{eqn:MMTheory}, 
we first perform computer simulations with variable number of beads representing the Rouse chain. 
The bead friction $\xi_0$, temperature $k_B T$ and statistical segment length $b$ are set to unity in the simulations without loss of generality, 
which sets the length($b$), time($\xi_0 b^2/k_B T$) and energy($k_B T$) scales. 
Direct simulations of eq.\ref{eqrouse} were performed using the predictor-corrector method\cite{Allen1989}. 
Detection of first-passage events was improved by computing the probability of unobserved events in each time step~\cite{OttingerBook}. 
Even if no crossing is observed before and after a time step, 
there is still a possibility that trajectory has crossed an interface during the step. 
If $q_1$ and $q_2$ are the distances between the interface and
the reaction coordinate before and after the time-step, such probability is
\begin
{equation}
P_{cross} = \exp(- \dfrac{q_1 q_2}{D \Delta t} ), \label{OttingerAlgorithm}
\end{equation}
where $D$ is the short time diffusion coefficient of the reactive coordinate, in our case, of the last bead: $D=k_B T/\xi_0$. A uniform random number $w$ on $[0..1]$ interval was generated and the unobserved event registered if $w<P_{cross}$. With this algorithm and the predictor-corrector integration we were able to use time step $\Delta t=0.05$, with smaller steps giving almost identical results. 

\begin{figure}[htbp]
\centering
\includegraphics[width=0.5\columnwidth]{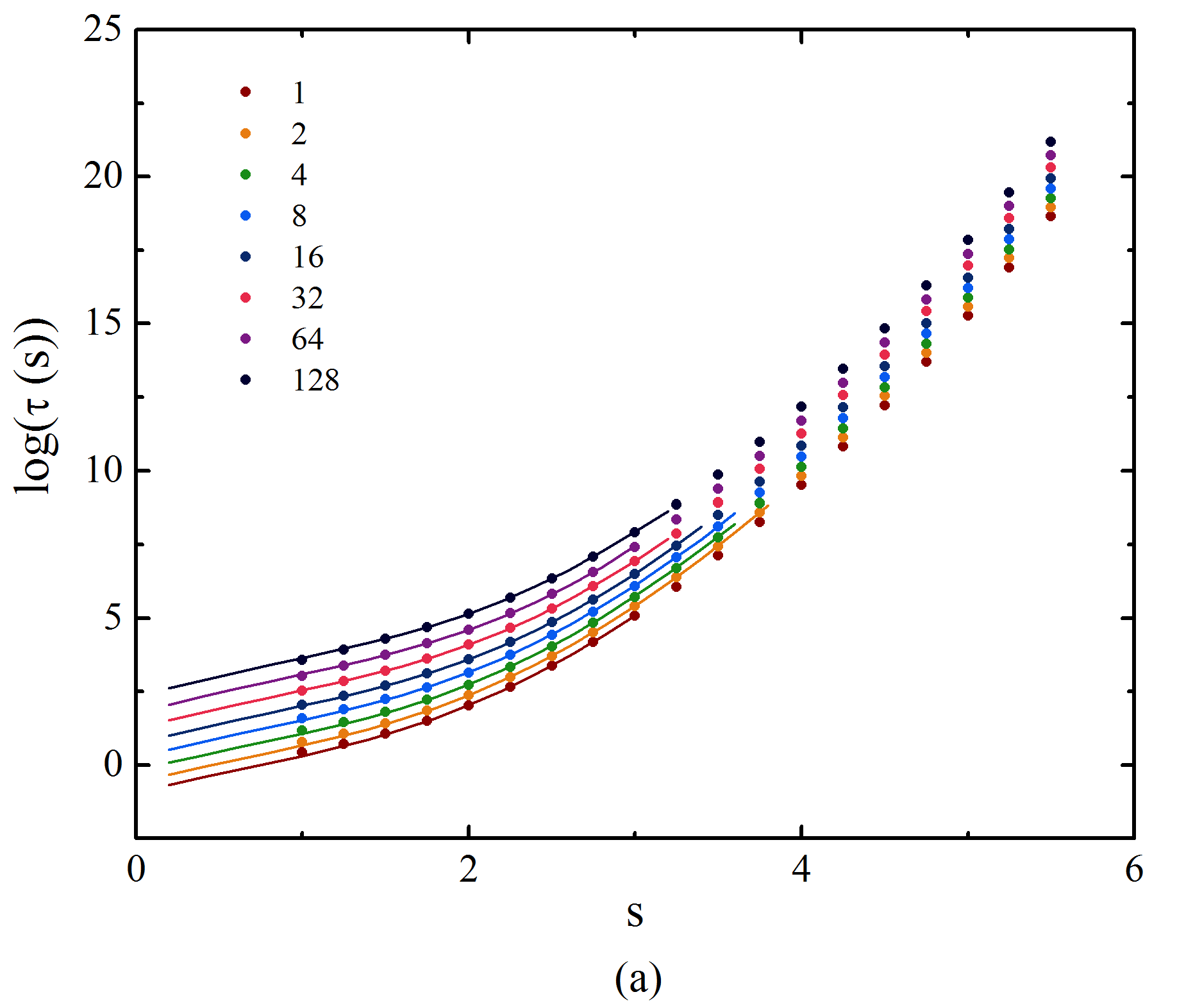}
\includegraphics[width=0.5\columnwidth]{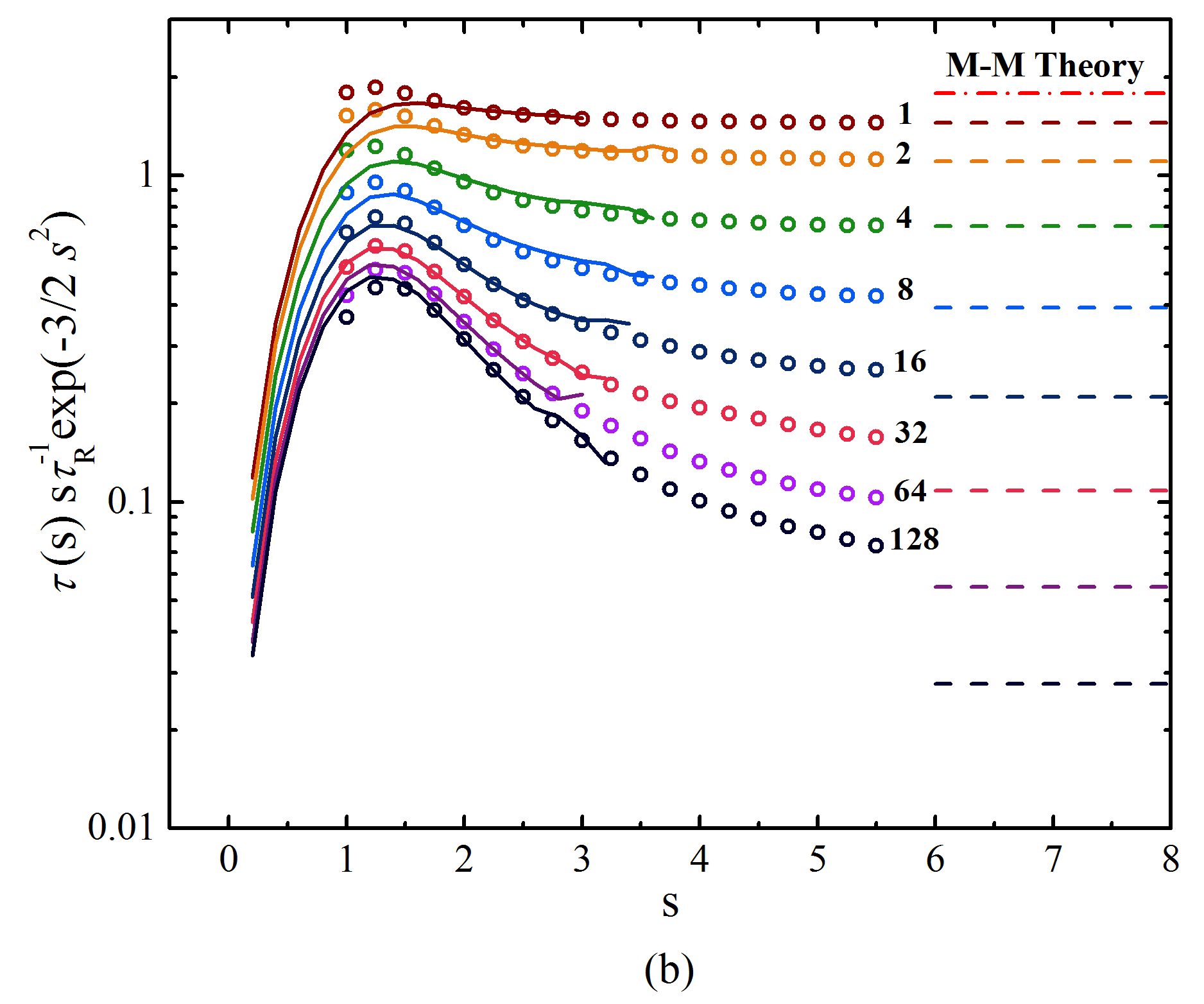}
\caption{(a) The logarithm of FP time $\tau(s)$ versus $s$ for FFS simulations (dots) and direct simulations (solid lines). (b) $\tau(s)s \tau_R^{-1} \exp(-3/2 s^2)$ versus $s$ for FFS simulations (circles) and direct simulations (solid lines),   the dashed lines are the prediction of Equation \ref{eqn:11_2}.  Milner-McLeish theory is shown by the red dash-dot line.} \label{ffs1dtube}
\end{figure}

Direct simulation results for the mean FP time are plotted in Fig.\ref{ffs1dtube} with solid lines for different numbers of beads N. The horizontal axis in both plots is the distance from the free chain end to the fixed end, $z$, normalized by the root-mean-square end-to-end distance that $s=z/(N^{1/2}b)$. 
In a continuously simulation, when the free end last crossing $s_0=0$ reaches $s>0$ for the first time, its time cost is recorded as the FP time for $s$. 
Fig.\ref{ffs1dtube}(a) shows decimal logarithm of the mean FP time. 
Clearly, the time grows very fast with $s$, approximately as $\exp(3s^2/2)$ as expected\cite{Doi1988}, and the simulations can only carry out the measurement up to $\tau(s)\approx 10^7$ or so. 
Fig.\ref{ffs1dtube}(b) shows the same data in reduced coordinates $\tau(s) s \tau_R^{-1} \exp(-3s^2/2)$. Here for clarity we divided by the trivial factor $\exp(U(s)/k_B T)=\exp(3s^2/2)$ predicted by all theories, and the typical fluctuation time $\tau_R$, and also multiplied by the expected prefactor scaling $s$ as suggested by eq.\ref{eqn:MMTheory}. Such a renormalized plot brings all data within one decade in vertical axis scale and allows clear comparison between theories and simulations. In particular, the Milner-McLeish result, eq.\ref{eqn:MMTheory}, is constant in this representation and is shown by dot-dashed line. Already from the direct simulations, it is clear that simulation results are significantly faster than the Milner-McLeish prediction.

\begin{figure}[htbp]
\includegraphics[width=0.8\columnwidth]{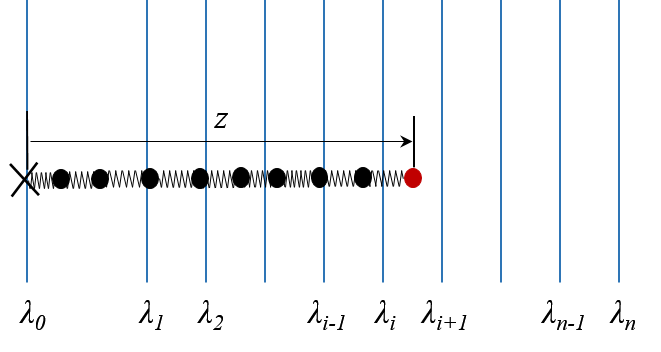}
\caption{The sketch of the direct and FFS simulation. The Rouse chain with one end fixed do one-dimensional Brownian motion. $z$ can be normalized into $s$ for the comparison between different chain length. The interface definition is only for FFS simulation.}\label{simulationSketch}
\end{figure}

In order to extend simulation results to longer times and facilitate detailed theory verification and calibration, we also performed forward-flux sampling(FFS) simulation of the same model. FFS~\cite{FFS2005} is a simple method to compute FP times of unlikely events by splitting the phase space of the system by $n+1$ interfaces defined such that the starting points of the trajectory are on the left  of $0$-th interface, 
and the final point --- on the last, $n$-th, interface, as shown in Fig.\ref{simulationSketch}. 
Besides, the interfaces must be defined such that every trajectory has to pass all interfaces in order to get to the reactive state. 
In our case, the position of the $i$-th interface can be simply defined by the position of the free end $R_N=\lambda_i$. We found that $\lambda_0=0$,  $\lambda_i= (1 + 0.25*(i-1))N^{1/2}b, \quad i\ge 1$ gives the most accurate results, avoiding systematic errors due to very small interface distance and large statistical errors due to large interface distance. 

The simulation then proceeds in two stages. In the first stage, we run one long simulation for time $T_{0}$ and count the number of crossings, $N_0$, of the first interface $\lambda_1$ by the trajectories which last crossed interface $\lambda_0$, rather than $\lambda_1$. 
These results are in the attempt frequency 
\begin{equation}
\nu_0=\frac{N_0}{T_0}
\end{equation}  
Besides counting the crossings, we store the full chain configurations at the moments of these crossings. 

In the second stage, we run many short consecutive simulations for interfaces $1$ to $n-1$ in sequence. For the interface $\lambda_i$, the simulation starts from the stored points on the interface $\lambda_i$ (selected at random from the database) and finish when they either reach the next interface $\lambda_{i+1}$(successful run), or go back to the $0$-th interface (unsuccessful run). The fraction of successful runs $N_i/M_i$ gives an estimate of the probability to progress from one layer to the next, $P(\lambda_{i+1}|\lambda_{i})$, where $M_i$ is total number of runs from layer $i$, and $N_i$ is the number of successful runs. 
Thus, the mean FP time is given by

\begin{equation}
\tau(\lambda_n) = \frac{1}{\nu_0 \prod_{i=1}^{n-1} P(\lambda_{i+1}|\lambda_{i})}. \label{eqFPT}
\end{equation} 

The value of $M_i$ has a decisive effect on the statistical error of the final outcome, with the best strategy to increase $M_i$ for higher energy barriers between the layers to ensure an approximately constant $N_i$. A simple way to determine $M_i$ is to run a few simulations with smaller $M_i$ and get the rough ratio of $P(\lambda_{i+1}|\lambda_{i})$, and estimate $M_i$ for an expected $N_i$. Ref.\cite{ffsInterface} recommends selecting interface distances such that $P(\lambda_{i+1}|\lambda_{i})> 0.3$. Our selection satisfies this criteria. By running a quick simulation for $N=1$, the proper $M_i$ can be obtained. Using the same $M_i$ and the same distance defined by $s$ for $N=1$, a proper ratio $P(\lambda_{i+1}|\lambda_{i})$ for larger $N$ is also guaranteed since the $P(\lambda_{i+1}|\lambda_{i})$ increases with larger $N$. 

The mean FP time $\tau(s)$ for various normalized extension length $s$ are presented for different chain lengths $N$ in Fig.\ref{ffs1dtube}. The FFS result is the harmonic average of FP times from independent runs. More discussion on the averaging method and the error can be found in Appendix A. Comparing with direct simulations, two methods disagree slightly for $s< 1.5$. In this region, the FP time given by FFS is inaccurate since the energy barrier is below $3.5k_BT$, different FP events are not independent from each other and therefore distribution of FP times is not single exponential. In the region of $s > 1.5$, two simulation methods are consistent with each other. FFS method is able to predict the FP time till $s=5.5$, and the chain length up to $N=128$. In the normalized plot in Fig.\ref{ffs1dtube}(b), all curves develop from a negative slope to a plateau after exceeding some transition length $s_t$.  The slopes of the curves from the peak to $s_t$ increases with increasing $N$. In the mean time, the transition length $s_t$ also increases. 
One finds that the result differs from the Milner-McLeish theory significantly. 
When increasing $N$, the FP time becomes much shorter then their prediction, leading to the difference of a factor of 10 at $s=3$ and $N=128$, and even bigger for larger $s$ and $N$. This shows conclusively that the one mode assumption of the Milner-McLeish theory is inadequate and better theory must be developed. Note that this discrepancy is much bigger than the 20\% reported by Vega et.al.~\cite{Vega:2002}. It suggests that the multi-dimensional character of the end bead evolution is essential, motivating the next two sections.

\section{Exact asymptotic theory}

In this section we step away from the one-mode approximation of the Milner-McLeish theory and apply exact asymptotic Kramers theory\cite{Kifer1974,Meerkov:1988,Hanggi:1990} to the full problem in $N$-dimensional Rouse modes space.
We first briefly recall the theory here.
Consider a Brownian particle with friction $\xi$ diffusing in
a multi-dimensional space of dimension $d$, subject to the potential $U(\bfm{R})$. 
In the over-damped regime, its probability density distribution $\psi(\bfm{R},t)$ obeys the Smoluchowski equation
\begin{equation}
\xi\frac{\partial\psi(\bfm{R},t)}{\partial t}=\nabla\Big( \psi(\bfm{R},t) \nabla
U(\bfm{R})  +k_B T\nabla\psi(\bfm{R},t)\Big)  \label{eqsmol}%
\end{equation}
which can be divided into two equations%
\begin{align*}
\frac{\partial\psi}{\partial t} &  =-\nabla J\\
J(\bfm{R},t) &  =-\frac{1}{\xi}\psi\nabla U-\frac{k_B T}{\xi}\nabla\psi
\end{align*}
where $J(\bfm{R},t)$ is the current at point $\bfm{R}$. 
The first equation is called
continuity equation and the second --- Fick's law.

Since the equilibrium solution of such equation is given by Boltzmann distribution
$\psi_{eq}(\bfm{R})\sim\exp\left(  -U(\bfm{R})/k_B T\right)  $, it is convenient to change the
unknown function to
\[
\phi(\bfm{R},t)\equiv\psi(\bfm{R},t)\exp(U(\bfm{R})/k_B T)
\]
such that $\phi(\bfm{R},t)$ becomes constant in equilibrium. Then eq.\ref{eqsmol}
becomes%
\[
\frac{\xi}{k_B T}\exp\left(  -\frac{U(\bfm{R})}{k_B T}\right)  \frac{\partial\phi
(\bfm{R},t)}{\partial t}=\nabla\left(  \exp\left(  -\frac{U(\bfm{R})}{k_B T}\right)
\nabla\phi(\bfm{R},t)\right)
\]
and the expression for the current%
\begin{equation}
J(\bfm{R},t)=-\frac{k_B T}{\xi}\exp\left(  -\frac{U(\bfm{R})}{k_B T}\right)  \nabla\phi
(\bfm{R},t)\label{cur2}%
\end{equation}

The Kramers method of finding the mean FP time of escape is to consider a
system where particles are injected at the origin and deleted from the system
when they reach absorbing surface $\Omega,$ and find a steady state solution
of such system. In this case the average escape time will be given by the
total number of particles divided by the rate of injection $j_{in}$:%

\begin{equation}
\tau=\frac{1}{j_{in}}\int\psi(\bfm{R})d\bfm{R}=\frac{1}{j_{in}}\int\phi(\bfm{R})\exp\left(
-\frac{U(\bfm{R})}{k_B T}\right)  d\bfm{R} \label{tau}%
\end{equation}
where integration is over the whole space available to particles and $\psi(\bfm{R})$
and $\phi(\bfm{R})$ without time dependence are steady state values.

An asymptotic result as $T\rightarrow0$ was obtained and proved rigorously by
Meerkov\cite{Meerkov:1988,Hanggi:1990}, based on earlier papers of Kifer\cite{Kifer1974}. Here we show a simple
intuitive derivation which is not available in the original papers.

The crucial assumptions are that as $T\rightarrow0$, $\phi(\bfm{R})$ becomes
constant everywhere apart from a thin layer near the absorbing interface, and in
this thin layer the current is perpendicular to the surface. Let's call $x$
the direction orthogonal to the interface and $\bfm{Y}=\{y_{i}\}$ --- all other
directions. Then in the layer near absorbing surface we can write%

\begin{equation}
\frac{\partial\phi}{\partial x}=-\frac{\xi J_{x}(\bfm{Y})}{k_B T}\exp\left(  \frac
{U(\bfm{R})}{k_B T}\right)  ;\qquad\frac{\partial\phi}{\partial y_{i}}=0\label{e3}%
\end{equation}
where $J_{x}(\bfm{Y})$ is the current near the absorbing surface. 
The solution can be written as%
\begin{equation}
\phi(x,\bfm{Y})=\frac{\xi J_{x}(\bfm{Y})}{k_B T}\int_{x}^{x_{s}}\exp(U(x^{\prime
},\bfm{Y})/k_B T)dx^{\prime}\approx\frac{\xi J_{x}(\bfm{Y})}{U_{x}^{\prime}(x_{s},\bfm{Y})}%
\exp(U(x_{s},\bfm{Y})/k_B T) \label{e4}%
\end{equation}
where $x_{s}$ is the value of $x$ and the surface. Because we want $\phi(x,\bfm{Y})$
to be constant far away from the surface, we require that%
\[
\phi(x,\bfm{Y})=\frac{\xi J_{x}(\bfm{Y})}{U_{x}^{\prime}(x_{s},\bfm{Y})}\exp(U(x_{s}%
,\bfm{Y})/k_B T)=C\quad\rightarrow\quad J_{x}(\bfm{Y})=C\frac{U_{x}^{\prime}(x_{s},\bfm{Y})}%
{\xi}\exp(-U(x_{s},\bfm{Y})/k_B T)
\]
The last equation is sufficient to calculate the total current through the
surface, which must be equal to the injection rate%
\begin{align}
j_{tot}=j_{in}=&\int J_{x}(x_s,\bfm{Y}) d\bfm{Y} \nonumber\\
=&\frac{C}{\xi}\int U_{x}^{\prime}(x_{s}%
,\bfm{Y})\exp\Big(-\frac{U(x_s,\bfm{Y})}{k_B T}\Big) d\bfm{Y} \label{jtot}%
\end{align}
Let $\bfm{E}$ to be the Hessian matrix of the potential $U(x,\bfm{Y})$ under the new rotated coordinates
and we partition $\bfm{E}$ as follows,
\begin{align}
\bfm{E} = 
\begin{pmatrix}
  E_{xx} & \bfm{A}^T  \\
  \bfm{A} & \bfm{E'}  
\end{pmatrix}
\end{align}
where $E_{xx}$ is the entry of $\bfm{E}$ at the first row and the first column corresponding to $x$-direction, 
$\bfm{A}$ is a column vector with $N-1$ elements and $\bfm{E'}$ is $(N-1)\times (N-1)$ square matrix.
Since the exponential term in eq.\ref{jtot} can be treated as the distribution of multi-variables $\bfm{Y}$ conditional on $x=x_s$,
this conditional distribution is multivariate normal $(\bfm{Y}\vert x=x_s)\sim \bfm{N}(0,\bfm{\Sigma)}$ 
where covariance matrix 
\begin{align}
\bfm{\Sigma}= \bfm{E'}- \bfm{A} E_{xx}^{-1} \bfm{A}^T
\end{align}
such that 
\begin{align}
\det(\bfm{E})=\det(E_{xx}) \det(\bfm{\Sigma}) \label{eqn:MatrixPartition}
\end{align}


Suppose the minimum of the potential on the absorbing surface is reached at $(x_s, \bfm{Y}_s)$.
The integral of eq.\ref{jtot} is dominated by the potential in the area around $(x_s, \bfm{Y}_s)$
and can be approximated as
\begin{equation}
j_{in}\approx \frac{CU_{x}^{\prime}(x_{s},\bfm{Y}_{s})}{\xi}\exp(-U(x_{s},\bfm{Y}_{s}%
)/k_B T)\frac{(2\pi k_B T)^{(d-1)/2}}{\sqrt{\mathrm{det}(\bfm{\Sigma})}} \label{jtot2}%
\end{equation}

The mean FP time is then the ratio of total number of particles
$N_{tot}$ to the total current. The major contribution to the number of
particles comes from the bottom of the potential, where $\phi(\bfm{R})$ is
approximately equal to constant $C$:%

\begin{equation}
N_{tot}=\int\phi(\bfm{R})\exp(-U(\bfm{R})/k_B T)d\bfm{R}\approx\frac{(2\pi k_B T)^{d/2}}{\sqrt
{\mathrm{det}(\Lambda_{ij})}}C \label{ntot}%
\end{equation}
where $\Lambda_{ij}$ is the Hessian of the potential computed at the minimum
of the potential in the original coordinates.

Combining eqs.\ref{jtot2} and \ref{ntot}, we arrive to the final expression,
similar to the one from Meerkov's paper%

\begin{equation}
\tau=\frac{N_{tot}}{j_{in}}\approx\frac{\xi}{U_{x}^{\prime}(x_{s},\bfm{Y}_{s}%
)}\sqrt{\frac{2\pi k_B T\mathrm{det}(\bfm{\Sigma})}{\mathrm{det}(\Lambda_{ij})}}%
\exp(U(x_{s},\bfm{Y}_{s})/k_B T) \label{tmany}%
\end{equation}
We note many similarities with the one-dimensional result. 
In the multi-dimensional version, the first derivative of the potential at the
arrival point $x_{s}$ is replaced by the first derivative along the normal to
the surface. The second derivative at the bottom is replaced by the determinant
of the Hessian, both being proportional to the volume available to the
particles at the bottom of the potential, or the volume of the area where
$U(\bfm{R})<k_B T$. The new term $\mathrm{det}(\bfm{\Sigma})$ has a meaning of channel width
available to the particles as they arrive to the most probable absorbing point
$(x_{s},\bfm{Y}_{s})$.

Now we will apply the result of eq.\ref{tmany} to $1$-D discrete Rouse chain.
If one end of the Rouse chain is attached to $R_0=0$,
the transformation to eigenmodes $X_p$(or Rouse modes) and back is given by
\begin{align}
X_p=\frac{1}{N+1/2} \sum_{i=1}^{N} R_i \sin\frac{\pi i (p-1/2)}{N+1/2}; \; R_i=2\sum_{p=1}^{N} X_p \sin{\frac{\pi i (p-1/2)}{N+1/2}}
\end{align}
Since one end is fixed at origin,
the end-to-end vector of the Rouse chain is just 
\begin{align}
R_N=\sum_{p=1}^N \alpha_p X_p, \label{eqn:ete}
\end{align}
where 
\begin{align}
\alpha_p= 2 \sin{\frac{\pi N (p-1/2)}{N+1/2}}. \label{eqn:alpha_p}
\end{align}
The equations of motion for Rouse modes $X_p$ are then,
\begin{align}
\frac{dX_p}{dt} &=-\beta_p X_p +f_p(t) ; \quad p=1,2,\cdots,N \\
\langle f_p(t) \rangle &=0, \; \langle f_p(t) f_q(t')\rangle =\frac{2k_B T}{\xi_p} \delta(t-t')\delta_{pq}
\end{align}
where $f_p(t)$ are the random white noise
and $\beta_p$ are defined as the ratio between the spring constant $\kappa_p$ and the friction coefficient $\xi_p$
which are
\begin{align}
\kappa_p =& \frac{24 k_B T (N+1/2)}{b^2} \sin^2{\frac{\pi (p-1/2)}{2(N+1/2)}} \label{eqn:kappa_p}\\
\xi_p=&2(N+1/2)\xi_0 \label{eqn:xi_p}
\end{align} 
such that
\begin{align}
\beta_p=& \frac{24 k_B T(N+1/2) }{\xi_p b^2} \sin^2{\frac{\pi (p-1/2)}{2(N+1/2)}} \label{eqn:beta_p}
\end{align}
As shown in eq.\ref{eqn:xi_p}, the values of $\xi_p$ are the same for all $p$,
therefore we denote them as $\xi_X=\xi_p,p=1,\cdots,N$ in the rest of the paper.
Here, we also introduce two useful sums used frequently for the Rouse chain in the rest of paper,
\begin{align}
\sum_{p=1}^{N} \alpha_p^2 =& 2N+1 \label{eqn:alpha_p2}\\
\sum_{p=1}^{N} \frac{\alpha_p^2}{\beta_p} =& \frac{2N(N+1/2)\xi_0 b^2}{3k_B T} \label{eqn:alpha_p2Two}
\end{align}

Because $\det(\Lambda_{ij})$ is independent of the rotation of the coordinates and the potential is harmonic in each dimension $U(\bfm{X})=\sum_p \frac{\xi_X \beta_p}{2} X_p^2$
where $\xi_X\beta_p=\kappa_p$ is the spring constant of the potential in the $p$-th dimension, 
so $\det(\Lambda_{ij})=\prod_p \xi_X\beta_p$.
Because the absorbing boundary has the expression $\sum_p \alpha_p X_p =z$, 
the direction of $x$ is parallel to the unit vector
$\{ q_p\}=\Big\{\frac{-\alpha_p}{\sqrt{\sum_{p'} \alpha_{p'}^2}}\Big\}$.
Since $\det(\Lambda_{ij})=\det(\bfm{E})$ at every point in the coordinate space and equals to the product of the fluctuation amplitude in $x$-direction $|E_{xx}|$ and $\det(\bfm{\Sigma})$ according to eq.\ref{eqn:MatrixPartition},
then $|E_{xx}|$ is equal to $\dfrac{\det(\Lambda_{ij})}{\det(\bfm{\Sigma})}$,
which is nothing but the inverse of the spring constant of the potential in the $x$-direction
so that 
\begin{align}
\dfrac{\det(\Lambda_{ij})}{\det(\bfm{\Sigma})}=\frac{1}{\sum_p\frac{q_p^2}{\xi_X\beta_p}} 
=\frac{\xi_X \sum_p \alpha_p^2}{\sum_p \frac{\alpha_p^2}{\beta_p}} \label{eqn:Exx}
\end{align}

The derivative of potential in $x$-direction at $\{x_s, \bfm{Y}_s\}$ is 
$\frac{\partial U}{\partial X_p} \cdot q_p=\sum_p \xi_X \beta_p X_p \frac{\alpha_p}{\sqrt{\sum_{p'} \alpha_{p'}^2}}$.
Since $\{x_s,\bfm{Y}_s\}$ is the location of the potential minimum on the absorbing boundary $\Big\{\frac{z \alpha_p}{\beta_p \sum_{p'}\frac{\alpha_{p'}^2}{\beta_{p'}}} \Big\}$, 
we have
\begin{align}
U'_x(x_s,\bfm{Y}_s) =& \sum_p \xi_X \beta_p \frac{\alpha_p}{\sqrt{\sum_{p'} \alpha_{p'}^2}} \frac{z \alpha_p}{\beta_p \sum_{p'}\frac{\alpha_{p'}^2}{\beta_{p'}}} \nonumber\\
=& \frac{\xi_X z \sqrt{\sum_p \alpha_p^2}}{\sum_p \frac{\alpha_p^2}{\beta_p}} \label{eqn:Uprime} 
\end{align}

Substituting eqs.\ref{eqn:alpha_p2}, \ref{eqn:alpha_p2Two}, \ref{eqn:Exx} and \ref{eqn:Uprime} into eq.\ref{tmany}, we get 
\begin{align}
\tau(z)
=& \frac{\xi_0 }{k_B T}\frac{\sqrt{\pi}}{3\sqrt{6}}\frac{N^{3/2} b^3}{z}  \exp\Big(\frac{U(x_s,\bfm{Y}_s)}{k_B T}\Big)  \nonumber \\
=& \sqrt{\frac{32 \pi}{3}}\sin^2\Big(\frac{\pi}{4(N+1/2)}\Big) \frac{N^{3/2}b}{z} \tau_R \exp\Big(\frac{U(x_s,\bfm{Y}_s)}{k_B T}\Big) \label{eqn:11_2}
\end{align}
where the Rouse time of the discrete chain is $\tau_R=\frac{1}{\beta_1}=\frac{\xi_0 b^2}{12k_B T}\sin^{-2}\frac{\pi}{4(N+1/2)}$.
These asymptotic results for different $N$ values are shown in Fig.\ref{ffs1dtube}(b) by dashed lines together with the ones obtained from FFS simulations.
In the limit of large $N$, the result in eq.\ref{eqn:11_2} can be simplified as 
\begin{align}
\tau(z) = \frac{\pi^{5/2}}{2\sqrt{6}} \frac{\tau_R}{N} \frac{\sqrt{N}b}{z} \exp\Big(\frac{3z^2}{2 Nb^2}\Big)
\end{align}
which is $\frac{2}{N}$ times smaller than the Milner-McLeish result, eq.\ref{eqn:MMTheory}. Interestingly, this factor is exactly the ratio of friction of one bead to the effective friction used by Milner and McLeish, which means that the asymptotic result can be obtained by solving one-dimensional problem with spring constant $k=3k_BT/(Nb^2)$ and the friction of one bead. This means that in the limit of zero temperature or very large extensions $z$ the friction of other beads must be neglected.

One of the most important conclusion is that this result can not be expressed as a function of reduced variable $s$ and $\tau/\tau_R$ only because it has an additional factor of $\frac{1}{N}$:
\begin{align}
\frac{\tau(s)}{\tau_R} = \frac{1}{N} \frac{\pi^{5/2}}{2\sqrt{6}} \frac{1}{s}  \exp\Big(\frac{3s^2}{2}\Big) \label{eqn:taus_asymp}
\end{align}
Thus, it becomes arbitrary small in the limit of continuous Rouse chain. This does not seem to be physical, and a more detailed theory of the next section will reveal the reasons behind such behavior.


\section{Projection onto minimal action path}

In mathematics, a widely known approach to FP problem is the Freidlin-Wentzell theory\cite{Freidlin1998},
which operates with the probability of the Brownian paths in terms of an action functional.
It uses the fact that the most probable path is given by minimizing the action functional associated with the system.
Sophisticated numerical techniques have been developed to find such transition path and transition rates in complex systems\cite{Hanggi:1990,Ionova:1993,Olender:1996,Olender:1997,Dellago:1998,EWN:2002,EWN:2007}.
In this section, we will apply this approach to our system. For simplicity, we first consider 1-D case.

\subsection{Minimal action path}

\subsubsection{1-dimensional case}

Suppose a particle is injected at $x=0$ and $t=0$ into a $1$-D system with harmonic potential $U(x)=\frac{1}{2}k x^2$, 
where the spring constant $k=\beta\xi$, $\xi$ is the friction coefficient of the particle.
This is one of the simplest stochastic process called Ornstein-Uhlenbeck process.
The particle fluctuates in the potential and  is deleted from the system at $x=z$ and $t=t^*$.
In order to identify the trajectory of the particle,  
a small time interval for recording the position of the particle is set to $\Delta t$
such that the trajectory consists of $n$ segments and corresponding $n+1$ coordinates where $n=t^*/\Delta t$.
In other words, 
the trajectory of the particle can be somehow treated as a ``bead-spring chain'' with $n+1$ beads 
where two ends fixed at $x=0$ and $x=z$, respectively. 

Since Ornstein-Uhlenbeck process is Markovian, the chain segments are independent from each other.
The probability of the chain having configuration $\{\bfm{r}\}$
is simply the product of the transition probability density associated with $P(r_{i+1}\vert r_{i})$,
\begin{equation*}
P(\{r_1,r_2,\cdots, r_{n+1}\})=P(r_{n+1}\vert r_{n})P(r_{n}\vert r_{n-1})\cdots P(r_2\vert r_1)P(r_1)
\end{equation*}
The transition probability for Ornstein-Uhlenbeck process is well known
\begin{equation*}
P(r_{i+1}\vert r_i)=\frac{1}{\sqrt{2\pi D(1-e^{-2\beta \Delta t})/\beta }}
\exp\Big( -\frac{\beta (r_{i+1}-r_{i}e^{-\beta \Delta t})^2}{2D(1-e^{-2\beta \Delta t})}\Big)
\end{equation*}
where $D=k_B T/\xi$. 
Then
\begin{equation}
P(\{\bfm{r}\})=\Big(\frac{\beta }{2\pi D(1-e^{-2\beta \Delta t})}\Big)^{n/2}
\exp\Big[-\frac{\beta}{2D(1-e^{-2\beta \Delta t})} \sum_{i=1}^{n}(r_{i+1}-r_{i}e^{-\beta \Delta t})^2\Big] \label{eqn:5}
\end{equation}
Once the number of segments $n$ and $t^*$ are chosen,
$\Delta t$ is constant such that the prefactor in front of the exponent term is just constant
which only contributes to the normalization of the probability.
Thus,
\begin{align}
P(\{\bfm{r}\})\sim & \exp\Big[-\frac{\beta}{2D(1-e^{-2\beta \Delta t})} \sum_{i=1}^{n}(r_{i+1}-r_{i}e^{-\beta \Delta t})^2\Big] \quad (if\; \Delta t \; is \; small) \nonumber \\
\sim & \exp\Big[-\frac{1}{4D} \sum_{i=1}^{n} \big(\frac{{\Delta r_i}^2  }{\Delta t} + 2r_i \Delta r_i\beta
+ r_i^2\beta^2\Delta t\big) \Big] \;\;\;\; (let \; \Delta t \rightarrow 0) \nonumber\\
=& \exp\Big[-\frac{1}{4D} \int \big(\dot{r}(t)+r(t)\beta\big)^2 dt\Big] \label{eqn:TarjProb}
\end{align}
where $\dot{r}(t)=\dfrac{dr}{dt}$.

According to eq.\ref{eqn:TarjProb},
finding the most probable trajectory $\{\bfm{r}\}$ is the same as 
finding the trajectory which minimizes action $A(\{r\})=\int \big(\dot{r}(t)+r(t)\beta\big)^2 dt$.
If we define the Lagrangian of the system as the integrand of the action $\bfm{L}(r,\dot{r},t)=\big(\dot{r}(t)+r(t)\beta\big)^2$,
the action will have an extremal if the Euler-Lagrange equation is satisfied along the trajectory
\begin{align}
\frac{d}{dt}\frac{\partial \bfm{L}}{\partial \dot{r}}=\frac{\partial \bfm{L}}{\partial r} \label{eqn:Lag}
\end{align}
Since the second derivative of the Lagrangian over space and momentum are both positive,
the action has a minimum along the trajectory.
Eq.\ref{eqn:Lag} leads to an ordinary differential equation for the minimum action trajectory
\begin{equation}
\ddot{r}(t) =\beta^2 r(t) \label{eqn:ddot}
\end{equation}
with the general solutions
\begin{equation}
r(t)=C_1 \exp(-\beta t)+C_2 \exp(\beta t) \label{eqn:11}
\end{equation}
where $C_1$ and $C_2$ are arbitrary constants.

In order to capture the correct optimal path,
the selection of a large enough time interval $[0,t^*]$ is essential\cite{EWN:2007,ZhouX:2008}.
If the absorbing boundary is far from the origin so that the energy differences between points on the boundary and origin are much larger than $k_B T$,
the optimal path is not sensitive to the value of $t^*$ as long as $t^*$ is reasonably large. 
To simplify the problem further using the reversibility of equilibrium dynamics,
we can assume that the particles are injected into the system at the absorbing boundary $x=z$ at $t=0$
and will eventually arrive in origin
such that $r(0)=z$ and $r(+\infty)=0$. 
Because the second term in eq.\ref{eqn:11} is divergent,
$r(t)$ will drop to origin if and only if $C_2=0$.
Therefore, the solution with the specific boundary conditions is 
\begin{equation}
M(t)=r(t) =z \exp(-\beta t) \label{eqn:x_pmin_1D}
\end{equation}
which is termed as minimal action path(MAP).
Note that the parameter $t$ here is not the real time but a convenient parametrization.

\subsubsection{$N$-dimensional case}

In $N$-dimensional case,
suppose the absorbing boundary is defined by the expression $\sum_p \alpha_p X_p =z$, where $p\in\{1,2,\cdots,N\}$ and $\alpha_p$ are real numbers.
To obtain MAP $\bfm{M}(t)$ we first notice that the action in multi-dimensional case is just a sum of actions in individual dimensions, and therefore eq.\ref{eqn:ddot} and its solution eq.\ref{eqn:x_pmin_1D} apply
\begin{equation}
  M_p(t)=C_p\exp(-\beta_p t)
\end{equation} 
To find the unknown coefficients $C_p$, we use two initial conditions. First, the starting point must lie on the absorbing plane
\begin{equation}
  \sum_{p=1}^N \alpha_p C_p = z
\end{equation} 
And second, near the absorbing plane the direction of MAP must be perpendicular to to absorbing plane. This is because on small length scales the dynamics is dominated by the thermal fluctuations and therefore potential can be neglected, which means that the particle will find the shortest path to the absorbing plane. Combining these two conditions leads to the solution
\begin{equation}
M_p(t)=\frac{z\alpha_p}{\beta_p \sum_{p'} \frac{\alpha_{p'}^2}{\beta_{p'}}}\exp(-\beta_p t)\label{eqn:x_pmin}
\end{equation}

It is easy to check that the arrival point $M_p(0)=(z\alpha_p/\beta_p)/\sum_{p'} \frac{\alpha_{p'}^2}{\beta_{p'}}$ corresponds to the minimum of the potential on the absorbing surface.

\begin{figure}[htbp]
\centering
\includegraphics[width=0.4\columnwidth]{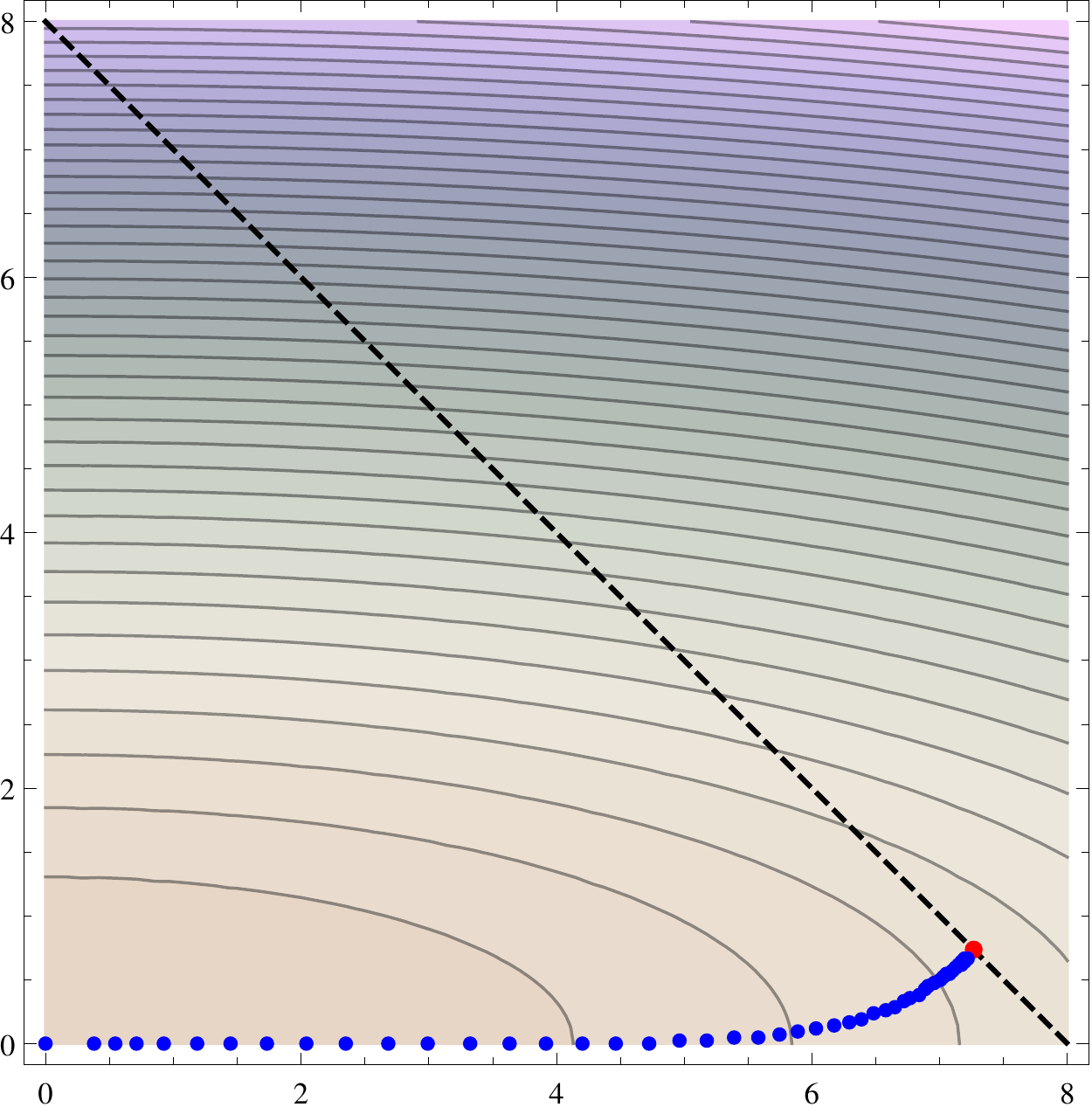}
\caption{MAP(eq.\ref{eqn:x_pmin}) between the origin and the location of minimal energy on the boundary in the system with potential $U(x,y)=x^2/2+10y^2/2$.} 
\label{fig:MAP}
\end{figure}

As an example, 
Fig.\ref{fig:MAP} shows the MAP between the origin and the location of minimal energy on the hyperplane $x+y=8$ in the system with potential $U(x,y)=x^2/2+10y^2/2$,
in which the curved lines are the contour plots of the potential. Note however that for smaller $z$ the actual minimum action path will deviate from our predictions since the action derived from eq.\ref{eqn:5} was computed without the effects of absorbing boundary.

The key idea of our method is to map the $N$-dimensional Kramers process onto a $1$-d problem along the MAP.
The effective potential along the MAP is defined as a function of the distance $l$ away from the origin along the path
such that the mean FP time for particles reaching the absorbing boundary can be derived from Kramers' solution eq.\ref{eqn:Kramers}.

Since we are interested in the FP time when the particle hits the absorbing boundary,
the trajectories should not go across the boundary.	
If the absorbing boundary is far away from the origin, 
the portion of the trajectories exceed the boundary is negligible.
However, if it is close to the origin,
the particles would find shorter path reaching the boundary
so that the MAP of eq.\ref{eqn:x_pmin} is not a good approximation of the most probable trajectory in the real system.

Without loss of generality,
we assume $\beta_i < \beta_j$ if $i<j$ so that the slowest mode is $X_1$.
In Kramers' solution, 
the inner integral would be dominated by the minimum of the potential for large $z$ values.
As shown in Fig.\ref{fig:MAP}, 
the MAP follows the $x$-axis near the origin which is the slowest mode 
such that the potential on the MAP near the origin is nothing but $\frac{1}{2}\beta_1 \xi_X l^2$
where $\beta_1 \xi_X$ is the smallest spring constant among all $\beta_p \xi_X$. 
Thus, using eq.\ref{eqn:Kramers}
\begin{align}
\tau(z) \approx \frac{\xi_X}{k_B T} \sqrt{\frac{2 \pi k_B T }{\beta_1\xi_X}}\int_0^{l(z)} \exp\Big(\frac{U(l)}{k_B T}\Big)dl \label{eqn:KramersSingle} 
\end{align}

Because the MAP is given by eq.\ref{eqn:x_pmin},
the distance $l(t)$ between the origin and $\bfm{M}(t)$ along the MAP can be obtained by
\begin{align}
l(t)=\int_{t}^{+\infty} \sqrt{\sum_p \Big(\frac{dM_p(t)}{dt}\Big)^2 } dt \label{eqn:lt}
\end{align}
and 
\begin{align}
\frac{dl}{dt}= \sqrt{\sum_p \Big(\frac{dM_p(t)}{dt}\Big)^2 } = |\dot{\bfm{M}}| \label{eqn:lt2}
\end{align}
Substituting eq.\ref{eqn:lt2} into eq.\ref{eqn:KramersSingle}, 
we have
\begin{align}
\tau(z) \approx \frac{\xi_X}{k_B T} \sqrt{\frac{2 \pi k_B T }{\beta_1\xi_X}}\int_0^{+\infty} \exp\Big(\frac{U(\bfm{M}(t))}{k_B T}\Big) |\dot{\bfm{M}}| dt \label{eqn:KramersSingle2} 
\end{align}

Clearly, MAP only gives the information about the most probable trajectory of the particles.
Along the MAP, the particles fluctuate in a relatively narrow channel with a certain width which depends on the potential landscape and the direction of the MAP. 
Since the MAP has a sharp turn approaching the absorbing boundary,
it is reasonable to believe that the width of the channel would also change when approaching the boundary.
We will investigate these effects in the next section.

\subsection{Conditional entropy}

In this section,
we will investigate how does the particle fluctuate perpendicular to the MAP. 
It is reasonable to believe that the distribution of particle positions in the direction perpendicular to the MAP is Gaussian
if the particle is far away from the absorbing boundary.
This assumption provides a simple approach for obtaining the entropy $S(t)$ of transverse fluctuation along the trajectory.
However, the distribution might deviate from Gaussian as the particle approaches the absorbing boundary.
To the first approximation, the density distribution of the particle is proportional to the equilibrium distribution in the absence of the absorbing boundary,
so that the entropy
\begin{align}
S(t)=-k_B \int P(\bfm{x}\vert \bfm{x}\in H(t))\ln P(\bfm{x}\vert \bfm{x}\in H(t)) d\bfm{x} \label{eqn:EntropyDef}
\end{align}
where $P(\bfm{x})\approx P_{eq}(\bfm{x})=\frac{1}{(2\pi)^{N/2} \prod_p\sqrt{k_B T/\beta_p\xi_X}}\exp\Big(-\sum_p\frac{\beta_p \xi_X x_p^2}{2 k_B T}\Big)$, $H(t)=\delta(\bfm{q}(t)\cdot (\bfm{x}-\bfm{M}(t)))$ is the hyperplane perpendicular to the MAP and passing through $\bfm{M}(t)$,
$\bfm{q}(t)=\frac{d\bfm{M}(t)}{dt}\Big(\frac{d l(t)}{dt}\Big)^{-1}$ is the unit tangent vector along the MAP
and $l(t)$ is the distance from the origin($t=+\infty$) along the MAP.

Thus,
\begin{align}
 P(\bfm{x}\vert \bfm{x}\in H(t)) = &
\frac{P_{eq}(\bfm{x}) \delta(\bfm{q}(t)\cdot (\bfm{x}-\bfm{M}(t)))}{ \int P_{eq}(\bfm{x}) \delta(\bfm{q}(t)\cdot (\bfm{x}-\bfm{M}(t))) d\bfm{x}} \label{eqn:ProbHyper}
\end{align}
Because the Dirac $\delta$-function can be represented by $\delta(\lambda)=\frac{1}{2\pi}\int_{-\infty}^{+\infty} e^{i\omega \lambda}d\omega$, 
the denominator in eq.\ref{eqn:ProbHyper} can be easily calculated as
$\Big(2\pi k_B T \sum_p \dfrac{ q_p(t)^2}{\beta_p\xi_X}\Big)^{-1/2} \exp\Big(-\dfrac{U(\bfm{M}(t))}{k_B T}\Big)$.
Substituting eqs.\ref{eqn:kappa_p}, \ref{eqn:xi_p}, \ref{eqn:beta_p}, \ref{eqn:ProbHyper} into eq.\ref{eqn:EntropyDef} gives
\begin{align}
S(t)=& k_B\Big(\sum_p \ln \sqrt{\frac{2 k_B T\pi e}{\beta_p \xi_X}} -\ln \sqrt{2 k_B T\pi e \sum_p\frac{q_p(t)^2}{\beta_p \xi_X}}\Big) \label{eqn:entropy1}
\end{align}
We can see that $\sum_p\frac{q_p(t)^2}{\beta_p \xi_X}$ in the second term in eq.\ref{eqn:entropy1} is nothing but the amplitude of particle fluctuation in the direction $\bfm{q}(t)$.
At $t=+\infty$, $q_p(t)$ are $0$ for all $p$ except $p=1$ since the MAP follows the slowest mode near the origin,
as illustrated in Fig.\ref{fig:MAP},
so that 
\begin{align}
S(+\infty)=k_B  \sum_{p\geq 2} \ln \sqrt{\frac{2 k_B T\pi e}{\beta_p \xi_X}}
\end{align}
which is exactly the entropy of $(N-1)$-dimensional multivariate normal distribution.
We are only interested in the entropy difference $\Delta S(t)$ between points on the MAP and the origin,
which is given by a very simple expression,
\begin{align}
\Delta S(t) = S(t) -S(+\infty) = \frac{k_B}{2} \ln \sum_p \frac{\beta_1 q_p(t)^2}{\beta_p}.
\end{align}

%
%

\subsection{Effective potential along MAP}

The effective potential along the MAP can be defined as
$U_{eff}(t)=U(\bfm{M}(t))-T\Delta S(t)$.
Substituting it into eq.\ref{eqn:KramersSingle2},
we get our central result for the mean first-passage time,
\begin{align}
\tau(z) =&
\sqrt{6\pi} \int_0^{+\infty} \frac{z}{\sqrt{N} b} \sqrt{F(N,t)} 
\exp\Big(\frac{3z^2}{2N b^2} F(N,t)\Big) dt
\label{eqn:tauS_entropyCorr}
\end{align} 
where 
\begin{align}
F(N,t) = \frac{N\xi_X b^2}{3k_B T \Big(\sum_p \frac{\alpha_p^2}{\beta_p}\Big)^2} \sum_p \frac{\alpha_p^2}{\beta_p} \exp(-2\beta_p t)
\end{align}
We note that $F(N,t)$ is related to the simple correlation function $\varphi(t)=\langle R(t)R(0)\rangle /\langle R^2 (t)\rangle$ with $R(t)=\sum_p \alpha_p X_p(t)$ without absorbing boundary by $F(N,t)=\varphi(2t)$. 
In addition,
we consider the entropy corrections at every point on the MAP instead of only at the absorbing boundary in section.3 and some other literatures\cite{EWN:2002},
which, as far as we known, is the first try.

\subsection{Discrete Rouse chain}

Now we will apply our theoretical prediction to the discrete Rouse chain
and $\tau(z)$ in eq.\ref{eqn:tauS_entropyCorr} is calculated using the parameters in eqs.\ref{eqn:alpha_p} and \ref{eqn:beta_p}.
After changing variable $t'=t/\tau_R$,
the mean FP time for the chain end is
\begin{align}
\tau(z)=\sqrt{6 \pi }  \tau_R
\int_0^{+\infty} 
\frac{z}{\sqrt{N}b} \sqrt{F(N,t')} \exp\Big(\frac{3z^2}{2Nb^2} F(N,t')\Big) dt'
\end{align}
where 
\begin{align}
F(N,t')
=&\frac{1}{2N(N+1/2)} \sum_{p=1}^{N} \frac{\sin^2(\frac{\pi N(p-1/2)}{N+1/2})}{\sin^2(\frac{\pi (p-1/2)}{2(N+1/2)})}
\exp\Big(-2 \big(\frac{\sin{\frac{\pi (p-1/2)}{2(N+1/2)}}}{\sin{\frac{\pi }{4(N+1/2)}}}\Big)^2  
t' \Big)
\end{align}

If $N\gg 1$, $F(N,t')$ has three different regimes as follows($t'$ is in the unit of $\tau_R$),
\[
F(N,t') = \left\{ 
  \begin{array}{l l}
    1- \frac{16 N}{\pi^2} t' , & \quad  t' \ll 1/N^2 \\
    1-  \sqrt{\frac{32t'}{\pi^3}} , & \quad  1/N^2 < t' < 1 \\
    \frac{8}{\pi^2} e^{-2t'} , & \quad  t' > 1
  \end{array} \right.
\]  
where the last regime is only contributed by the slowest mode and details of the derivation can be found in Ref.\cite{Doi1988}.

\begin{figure}[htbp]
\begin{center}
\includegraphics[width=0.45\columnwidth]{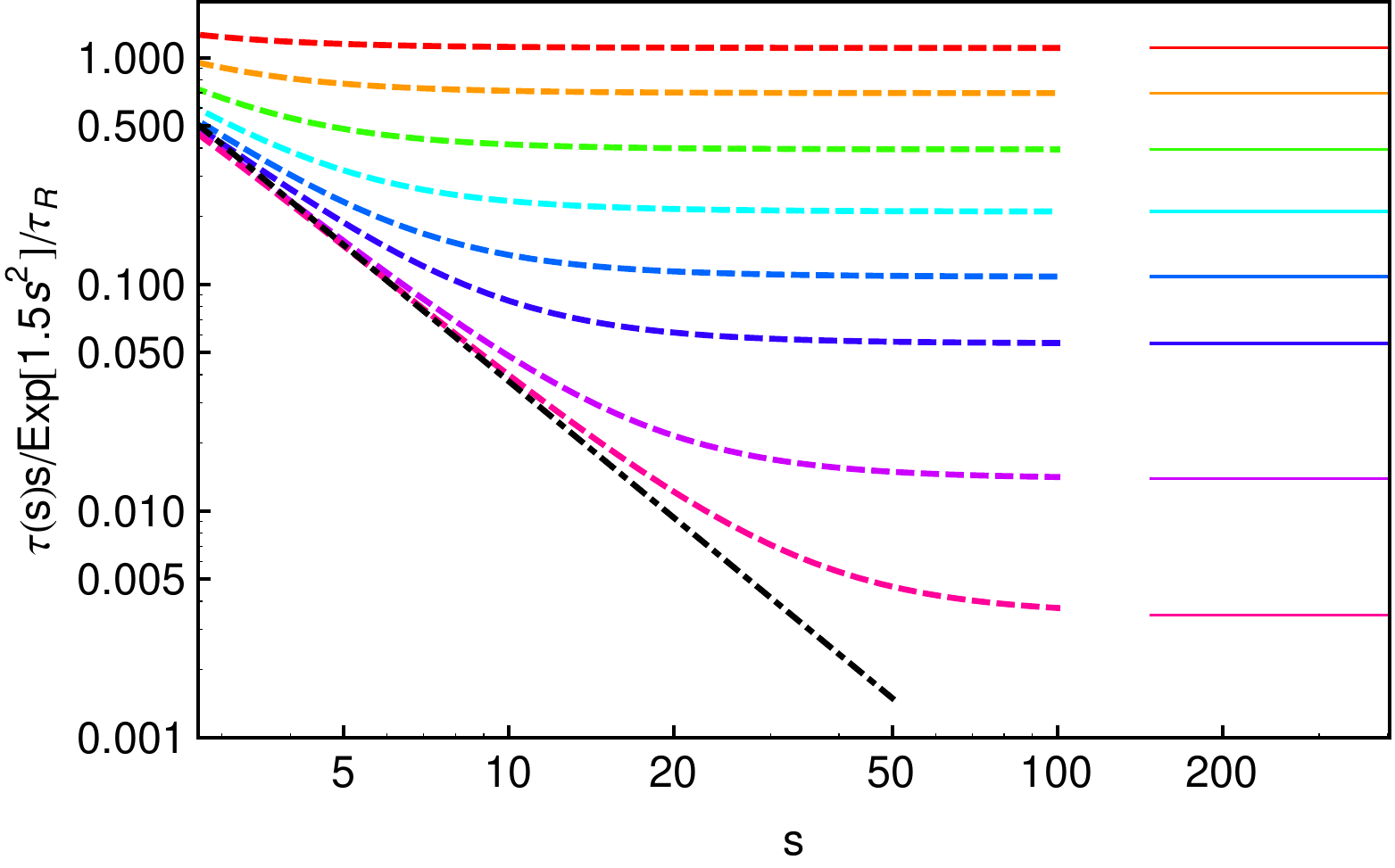}
\caption{Normalized FP time of the finite Rouse chain with different $N$ predicted by MAP projection theory (dashed lines from top to bottom correspond to $N=$2,4,8,16,32,64,256,1024). The asymptotic theory results from section $3$ are shown by solid lines.}
\label{fig:tauS_ContRouse2}
\end{center}
\end{figure}

For the reduced distance $s=\frac{z}{\sqrt{N}b}$ used in the previous parts of the paper,
the result further simplifies to
\begin{align}
\tau(s) = \sqrt{6 \pi }  \tau_R s
\int_0^{+\infty}  \sqrt{F(N,t')} \exp\Big(\frac{3s^2}{2} F(N,t')\Big) dt' \label{eqn:tauS2}
\end{align}
Fig.\ref{fig:tauS_ContRouse2} shows normalized mean FP time $\tau(s)$,
similar to the ones in Fig.\ref{ffs1dtube}(b),
for different number of Rouse beads $N$.
Clearly, we observe two different scaling regimes and the results at large $s$ values coincide with the prediction of the asymptotic theory in section 3. 
With increasing $N$,
the range of the first regime increases. 

\begin{figure}[htbp]
\begin{center}
\includegraphics[width=0.45\columnwidth]{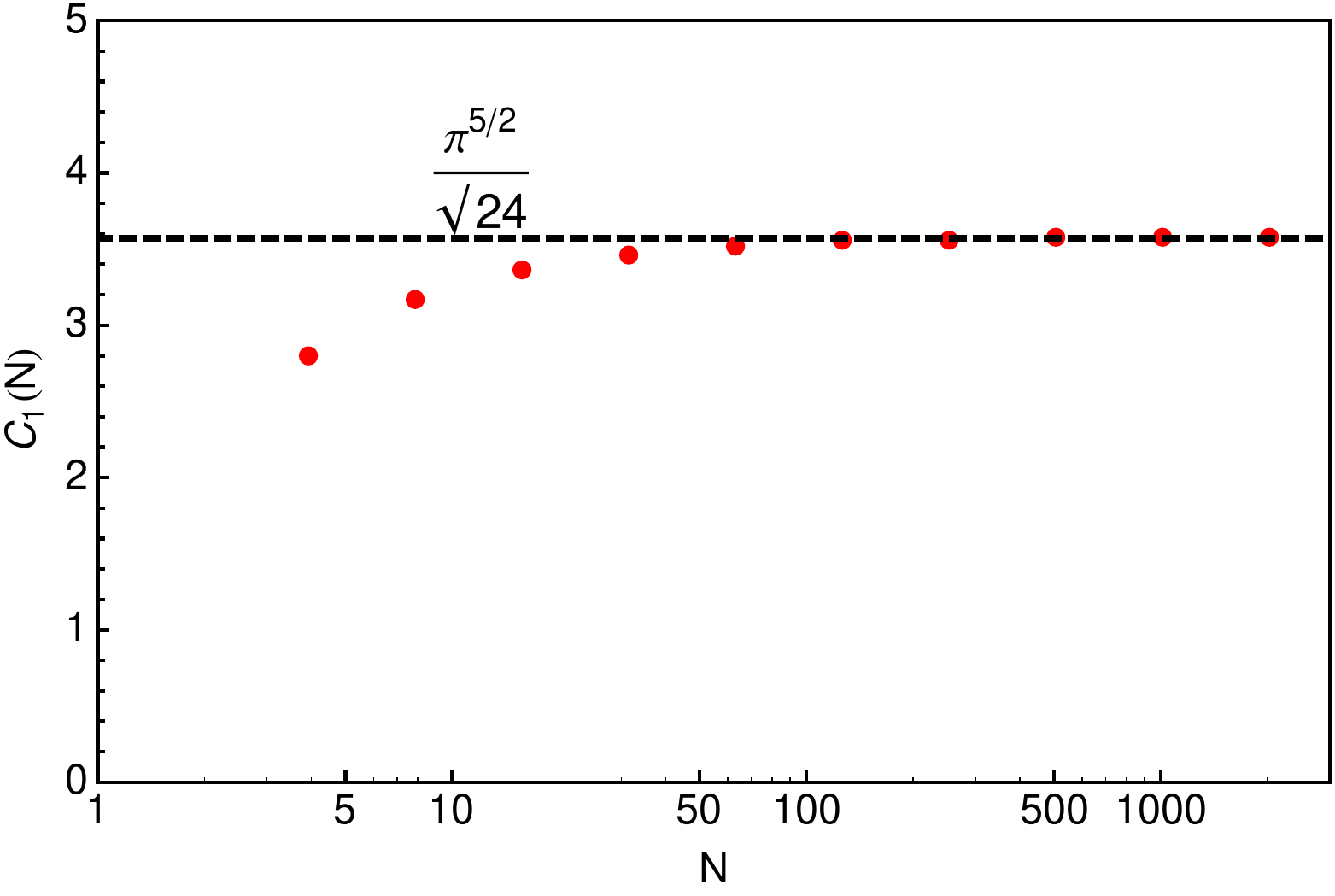}
\caption{$C_1(N)$ as a function of number of Rouse modes $N$.}
\label{fig:C1N}
\end{center}
\end{figure}

The integral in eq.\ref{eqn:tauS2} for large $s$ is dominated by small $t$ values 
such that
\begin{align}
\tau(s)\approx & \sqrt{6\pi } \tau_R s \sqrt{F(N,0)} 
\int_0^{+\infty} \exp\Big(\frac{3 s^2}{2 }\big(F(N,0)+F'(N,0) t \big)\Big) dt \nonumber \\
=& \frac{C_1(N)}{N s} \tau_R \exp\Big(\frac{3s^2}{2}\Big) ; \;\;\;\;\;\;\;\;\;\; C_1(N)=\sqrt{\frac{32\pi }{3}} N^2 \sin^2(\frac{\pi}{4(N+1/2)})  \label{eqn:tauS4}
\end{align}
We can see that
eq.\ref{eqn:tauS4} is the same as eq.\ref{eqn:11_2}
so that our theory gives the correct asymptotic results of $\tau(s)$ at large $s$ values in the systems with finite Rouse modes.
Fig.\ref{fig:C1N} shows $C_1(N)$ as a function of number of Rouse modes $N$.
If $N$ is large, $C_1(N)$ goes to a constant $\frac{\pi^{5/2}}{2\sqrt{6}}\approx 3.57$.

In the first regime, 
the integral in eq.\ref{eqn:tauS2} is dominated by $F(N,t)$ in the range $1/N^2\!<\! t\!<\!1$.
Thus, $\tau(s)$ can be approximated as following,
\begin{align}
\tau(s) \approx &\sqrt{6\pi } \tau_R s 
\int_0^{+\infty} \exp\Big(\frac{3 s^2}{2 }\big(1-\sqrt{\frac{32 t}{\pi^3}} \big)\Big) dt \nonumber \\
\approx & \frac{\sqrt{6} \pi^{7/2} \tau_R}{36 s^3} \exp(3s^2/2) \label{eqn:tauS_late}
\end{align}
which is illustrated by black dot-dashed line in Fig.\ref{fig:tauS_ContRouse2} showing the scaling of $s^{-3}$ in the prefactor.

After comparing eq.\ref{eqn:tauS4} and eq.\ref{eqn:tauS_late},
we see that the transition point of two scaling regimes is around $s^*\approx \sqrt{N}$.
In other words, the transition happens at $z^*\approx Nb$ 
where $b$ is the statistical segment length.
Thus, the prefactor of $\tau(s)$ will have $s^{-3}$ scaling until the chain reaches its full extension
and then it will go back to $s^{-1}$ scaling behavior.

A simple scaling argument can be used to explain such behavior. The Rouse chain under stretching
can be treated as a sequence of Pincus blob in the stretching direction, such that
the structure of the chain inside the Pincus blob is not significantly perturbed by the stretching force.
The asymptotic theory of section 3, eq.\ref{eqn:taus_asymp}, predicts that the FP time should scale as $\tau_R/(Ns) \exp\Big( 3/2s^2\Big)$, but this formula is only valid in the limit of fully stretched chains, $s>\sqrt{N}$. If we chose the level of coarsegraining to be the Pincus blob, we will satisfy this condition. In other words, to be physically meaningful, equation \ref{eqn:taus_asymp} should have the number of Pincus blobs instead of $N$ in the prefactor. Thus,  at particular $z$, the number of beads one must represent the Rouse chain with should be equal to the number of Pincus stretching blobs. This number is easily computed from the condition that the sizes of all blobs, each containing $g$ monomers, must add up to $z$, $\frac{N}{g} \sqrt{g}b=z$, thus giving the number of blobs 
\begin{equation}
N_{blobs}=\frac{N}{g}=\frac{z^2}{Nb^2}=s^2
\end{equation}
Using $N_{blobs}$ in eq.\ref{eqn:taus_asymp} instead of $N$ changes the scaling of $\tau(s)$ to $s^{-3}$. This argument also explains the transition to asymptotic regime $\tau(s) \sim s^{-1}$, which occurs when the number of Pincus blobs reaches $N$, giving $s^* \sim \sqrt{N}$ as discussed above.

\subsection{Comparison with FFS}

\begin{figure}[htbp]
\begin{center}
\includegraphics[width=0.5\columnwidth]{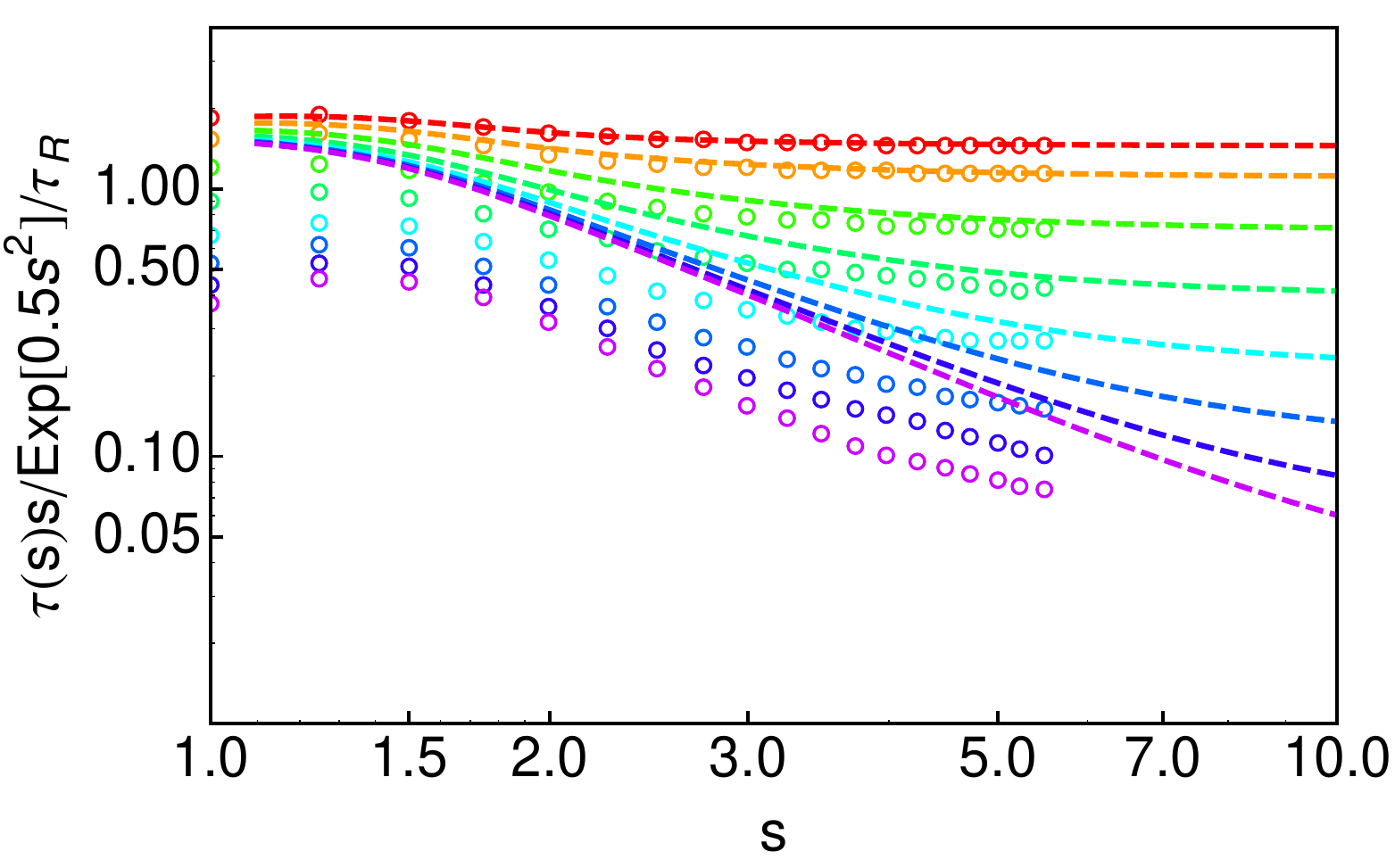}
\caption{Comparison of $\tau(s)s/\exp(1.5s^2)/\tau_R$ between direct FFS simulations(circles) and numerical calculation based on eq.\ref{eqn:tauS_entropyCorr}(lines) with different number of beads($N=1,2,4,8,16,32,64$ and $128$ from top to bottom). }
\label{fig:tauS_FiniteRouse3}
\end{center}
\end{figure}

The comparison of $\tau(s)$ between Forward-Flux sampling simulations(symbols) and our theoretical predictions(lines) is shown in Fig.\ref{fig:tauS_FiniteRouse3}.
The number of Rouse beads $N$ used in the simulations are $1,2,4,8,16,32,64$ and $128$ from top to the bottom. 
The agreement between results with $N\!\leq\! 4$ is reasonably good.
With $N=1$, our theory gives the same result as conventional 1-D Kramers' solution
so that the prediction from our theory lays on top of the simulation results exactly.
However, with increasing number of beads,
the agreement is getting worse in the regime we are interested in.
In such regime, our theory gives a correct scaling with $s$, but with the incorrect prefactor
which probably due to the fact that the process is not Markovian anymore after we project the trajectories onto the MAP.
In the following section,
we will provide an empirical expression to overcome the defect in the theory. 

\subsection{Combining analytical and simulation results}

Since we know both the exact asymptotic solution of $\tau(s)$ at large $s$ values 
and the one in the intermediate regime,
an empirical expression 
\begin{align}
\tau(s)=\Big(\frac{C_1(N)}{Ns}+\frac{C_2(N)}{s^3}\Big)\tau_R \exp\Big(\frac{3s^2}{2}\Big) \label{eqn:TauSFit}
\end{align}
can be proposed to recover the results in both regimes and to approximate the integral in eq.\ref{eqn:tauS2},
where $C_1(N)$ is given by eq.\ref{eqn:tauS4}.

\begin{figure}[htb]
\centering
\mbox{\subfigure{\includegraphics[trim = 0mm 0mm 0mm 0mm,width=0.39\columnwidth]{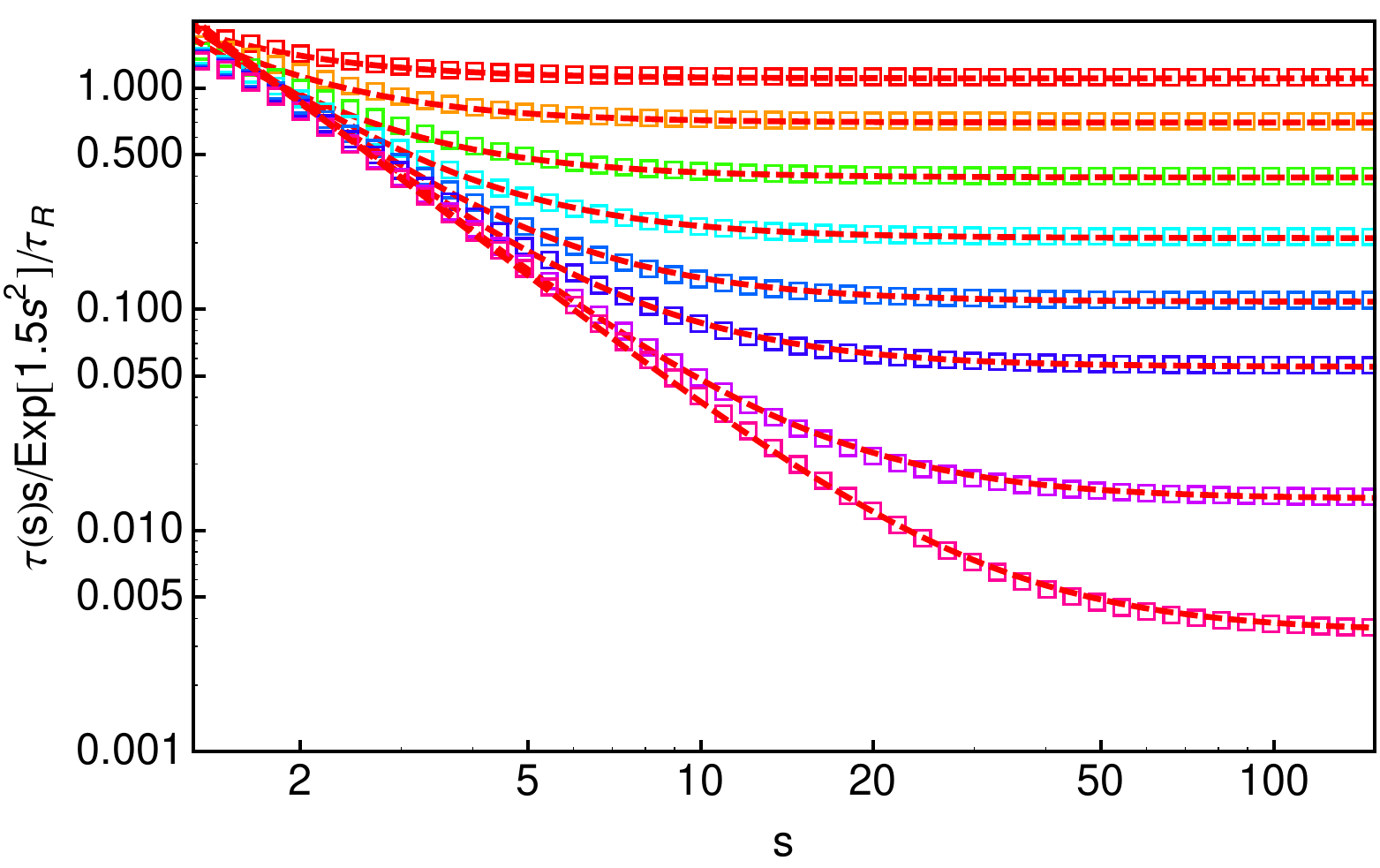}
\quad
\subfigure{\includegraphics[trim = -10mm 0mm 0mm 0mm,width=0.4\columnwidth]{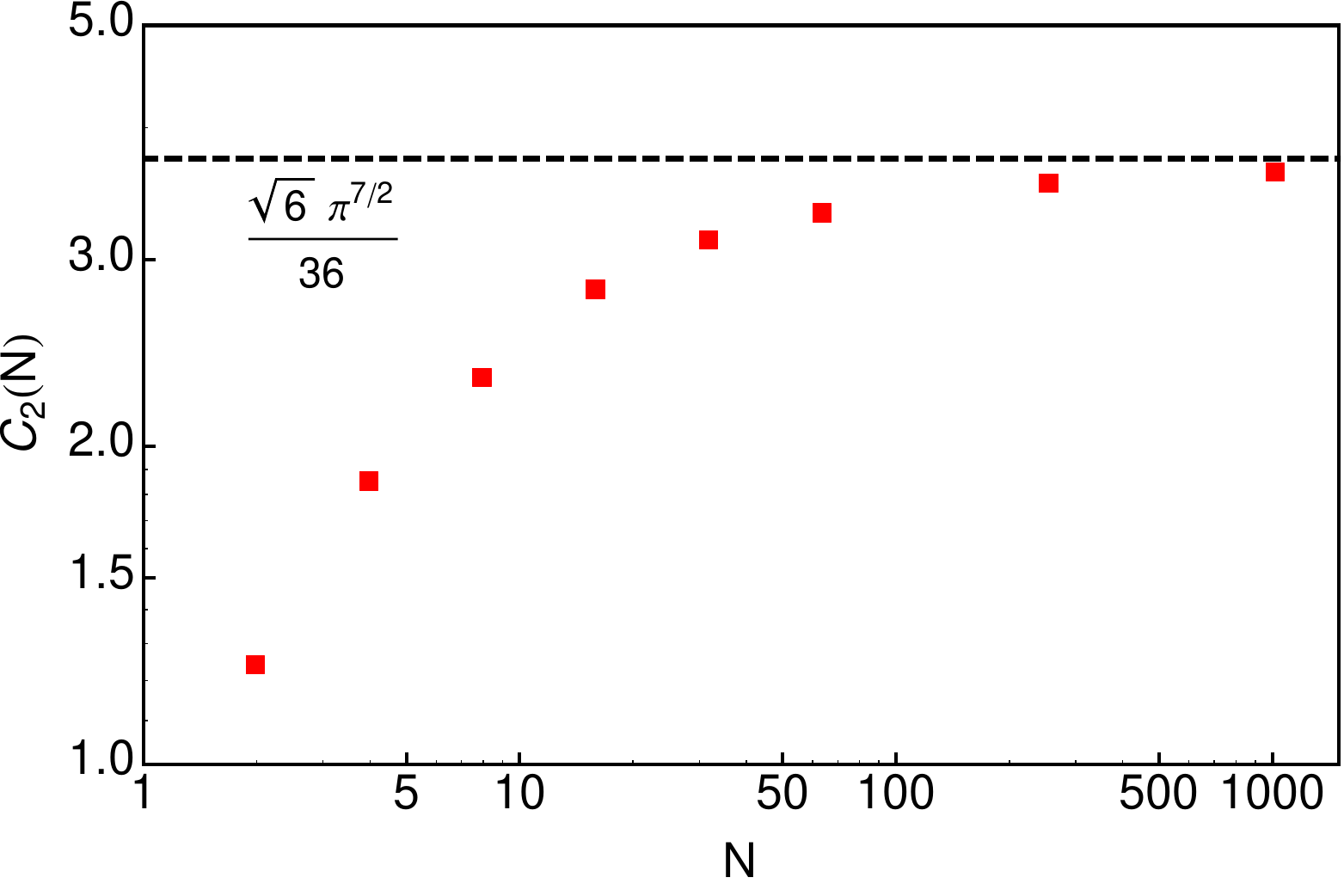} }}}
\caption{(a) Empirical fitting(lines) of theoretical prediction of $\tau(s)$(symbols) with different $N$ values($N=2,4,8,16,32,64,256$ and $1024$ from top to bottom). 
(b) Fitting parameter $C_2(N)$ as a function of $N$.}
\label{fig:EmpiricalTheory}
\end{figure}

Fig.\ref{fig:EmpiricalTheory}(a) shows the fitting of our theoretical calculations by using eq.\ref{eqn:TauSFit},
where $C_2(N)$ is a fitting parameter, and $C_1(N)$ is fixed by eq.~\ref{eqn:tauS4}. 
The fitting parameter $C_2(N)$ shown in Fig.\ref{fig:EmpiricalTheory}(b) increases with increasing $N$ 
and eventually reaches a plateau.
The relative fitting error,
defined as the maximum of $\dfrac{\tau_{exact}-\tau_{approx}}{\tau_{exact}}$ for each curve,
is smaller than $5\%$.

\begin{figure}[htb]
\centering
\mbox{\subfigure{\includegraphics[trim = 0mm 0mm 0mm 0mm,width=0.4\columnwidth]{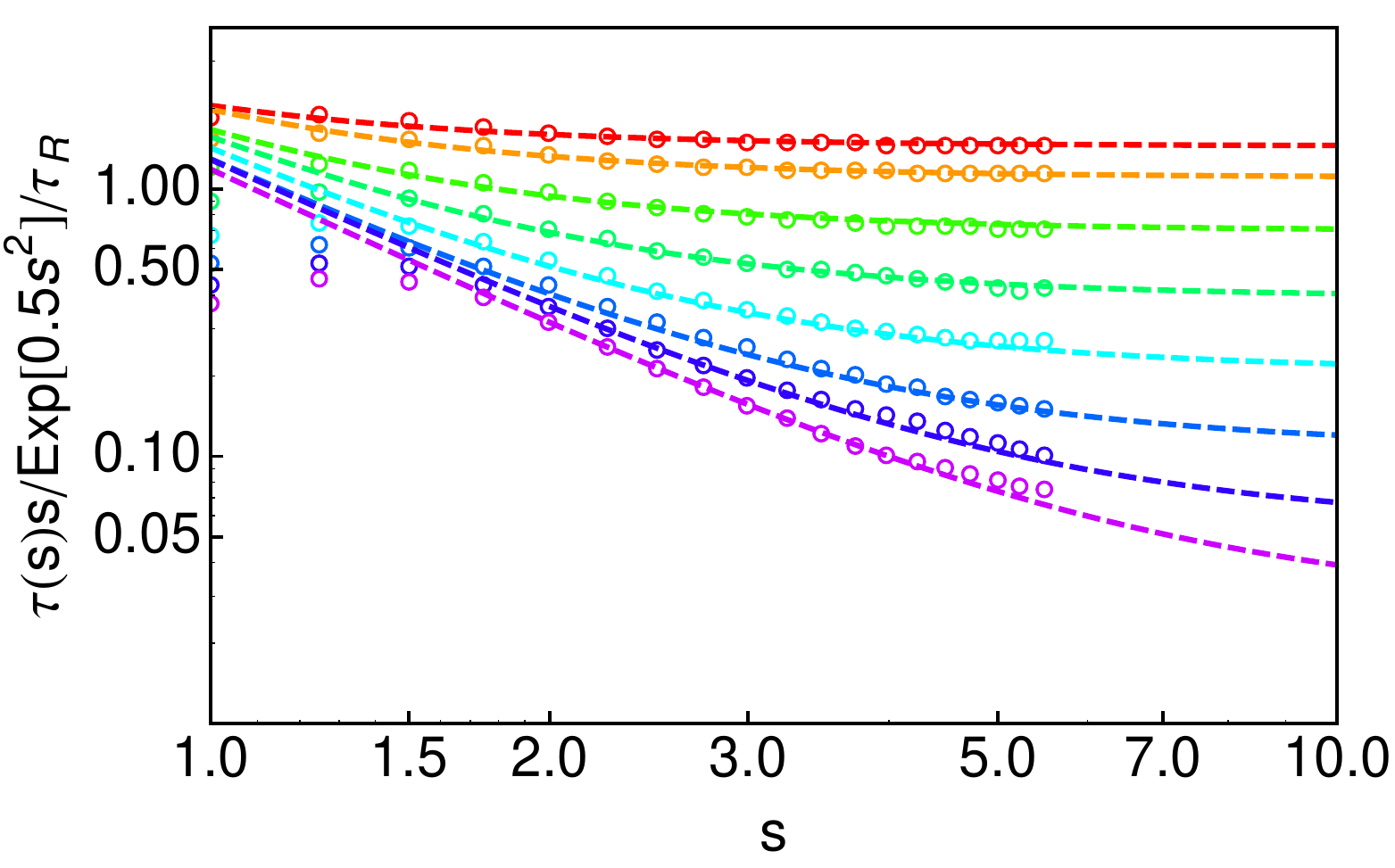}
\quad
\subfigure{\includegraphics[trim = -10mm 0mm 0mm 0mm,width=0.42\columnwidth]{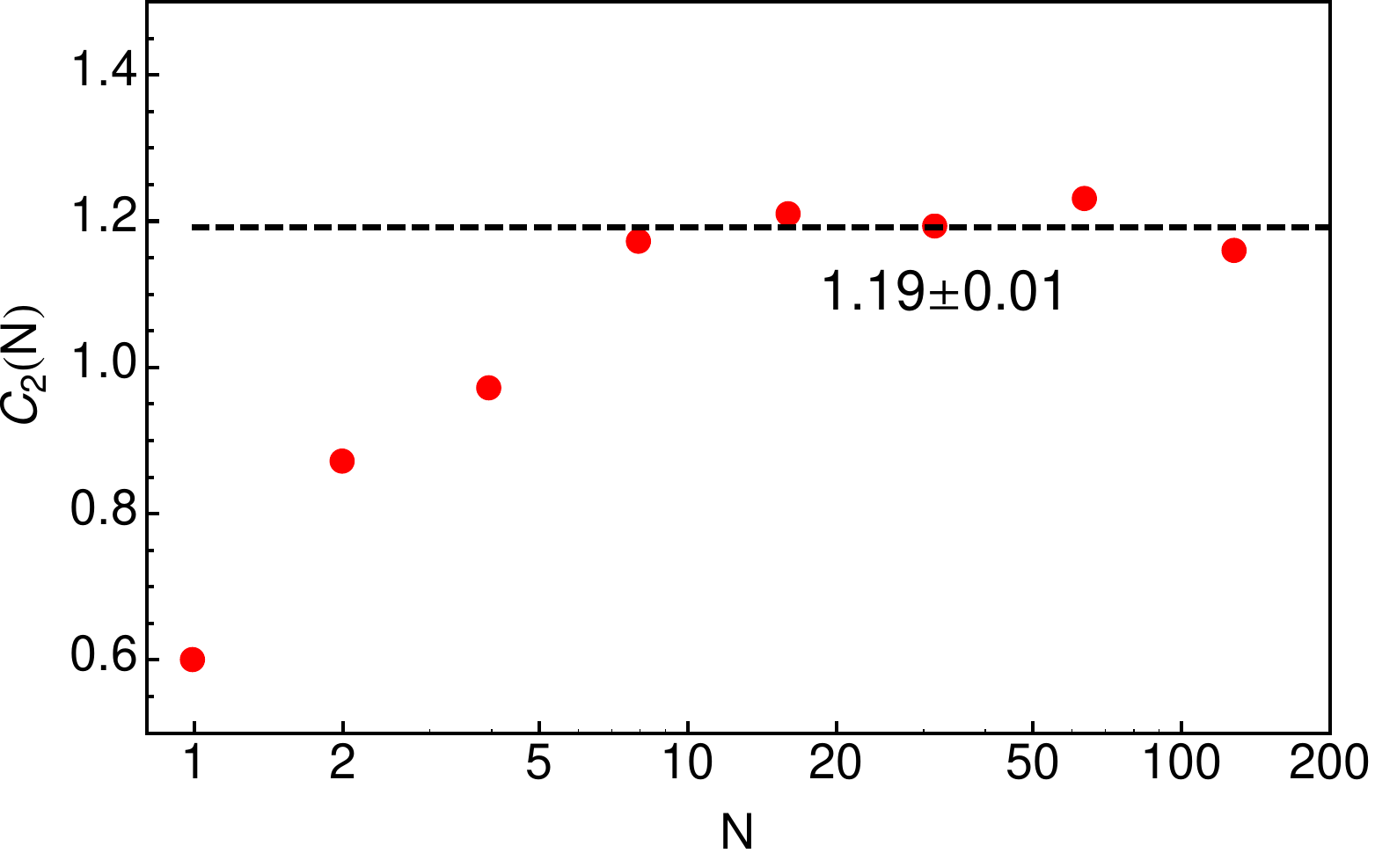} }}}
\caption{(a) Empirical fitting(lines) of $\tau(s)$ from FFS(symbols) with different $N$ values($N=1,2,4,8,16,32,64$ and $128$ from top to bottom). 
(b) Fitting parameter $C_2(N)$ as a function of $N$.}
\label{fig:EmpiricalSim}
\end{figure}

Fig.\ref{fig:EmpiricalSim}(a) shows the fitting of FFS results using eq.\ref{eqn:TauSFit}
where $C_2(N)$ is again the fitting parameter.
The fit is extremely good for all chain lengths and for $s>1.5$. For smaller $s$ neither Kramers' theory, nor FFS method are applicable.
Fig.\ref{fig:EmpiricalSim}(b) shows that the fitting parameter $C_2(N)$ reaches a plateau at smaller $N$ values comparing to the one in the theoretical calculations.
The value of $C_2(N)$ at large $N$ is about $3$ times smaller than the theory predictions. 
Thus, a combination of theory and FFS simulation leads to a simple analytical expression for the FP time valid for $N>10$,
\begin{align}
\tau(s) =\tau_R \left(\frac{3.57}{N s}+\frac{1.19}{s^3}\right)\exp\Big(\frac{3s^2}{2}\Big)
\end{align}

\section{Summary and Conclusions}

In this work, 
we have studied the first-passage problem for the chain end of discrete and continuous 1-D Rouse chains, 
as a proxy for dynamics of arm-retraction of isolated star polymers in a network. 
In the widely known Milner-McLeish theory, 
star arms are represented by Rouse chains inside their confining tubes 
and further replaced by one bead attached to the branch point by a harmonic spring. 
The mean disengagement time of a tube segment $\tau(z)$ is nothing but the first-passage time of arm-end reaching it, 
which can be calculated through Kramers' solution $\tau(z)\sim z^{-1}\exp(U(z)/k_B T)$. 

In order to check the validity of the Milner-McLeish theory, 
we carried out direct simulations to collect the first-passage time of Rouse chain extension
and found that the mean FP time drops significantly 
if the arm is represented by Rouse chain with more beads instead of a single bead.
Since the large deviations of the Rouse chain happen rarely,
an efficient simulation method called forward-flux sampling was applied to obtain results for large extensions.
The results from FFS simulations show that the prefactor of mean FP time $\tau(z)$  has $z^{-1}$ scaling only for very large extensions, 
but different scaling behaviour $\tau(z) \sim z^{-3}$ in the intermediate regime.
We argue that the large deviations of Rouse chain should be considered as a multi-dimensional Kramers' problem.

In order to solve the multi-dimensional Kramers' problem,
we first apply an asymptotic theory, valid in the limit of zero temperature. The solution has many similarities with the one-dimensional case.
We found that the theory becomes valid only for very large extensions $z\sim Nb$, corresponding to a fully extended chain, and fails in the physically relevant regime $\sqrt{N}b < z <Nb$. 

To describe this regime of moderate extensions,
a new theory was proposed, which involves projecting the dynamics onto the most probable trajectory between the origin and the absorbing boundary, termed ``minimal action path'' (MAP). We found that in case of Gaussian processes, MAP has a simple analytical expression. In addition,
we calculated the fluctuation perpendicular to the MAP at every point on the path
and added such entropic correction to the effective potential along the MAP.
Then, the mean FP time can be derived based on the conventional Kramers' solution along the MAP.
Our theory predicts that the prefactor of mean FP time has $z^{-3}$ scaling before the discrete Rouse chain reaches its fully extension and $z^{-1}$ otherwise. 
For the continuous Rouse chain,
the scaling is always $z^{-3}$ since the chain is never fully extended. 
Despite of more detailed treatment and correct scaling predictions, the constant prefactor in our theory for the first regime is about 3 times larger than the one obtained from the FFS simulations. This is probably caused by the projection of the $N$-dimensional system onto a one-dimensional reaction coordinate, which contains a hidden Markovian approximation. This hypothesis is supported by the fact that our solution for continuous chain, eq.\ref{eqn:tauS_late}, is identical to the result of ref.\cite{Marques:2001}, although it used very different method and the integral expression in ref.\cite{Marques:2001} is different from our result. Since the method used in ref.\cite{Marques:2001} was later shown to be suffering from a hidden Markovian approximation\cite{Likthman2006}, we suspect that our new method did not manage to avoid similar approximation. 

To rectify the disagreement, we combined simulation and analytical results into a simple formula for the mean FP time of discrete and continuous Rouse chains for large extension. In the future, the analysis of mean FP time of star polymers based on real systems will be carried out 
and the theory will be examined\cite{Cao:2015c}.

\appendix

\section{Statistical errors in the Forward-Flux sampling simulations}

Because FFS method works in a propagative manner, the simulation on each interface cannot be carried out simultaneously, which restricts parallelization. Meanwhile, it is expensive to explore the phase space especially in high-dimensional cases. For example, for the case of chain length $N=128$, the time that the chain evolves from one conformation to another uncorrelated conformation is $\tau_R \approx 2230$. The two conformations represent only two separate regions in a broad phase space, which is far from being enough to provide a good distribution on the first surface $\lambda_1$. In order to improve the accuracy with reasonable computational cost, one usually conducts a series of FFS simulations independently from different starting conformations, and calculate the FP time by averaging. This idea is equivalent to exploring the phase space from scattered points, although single FFS simulation can only explore locally, more simulations will be able to sample the whole space.

\begin{figure}[htbp]
\centering
\includegraphics[width=0.45\columnwidth]{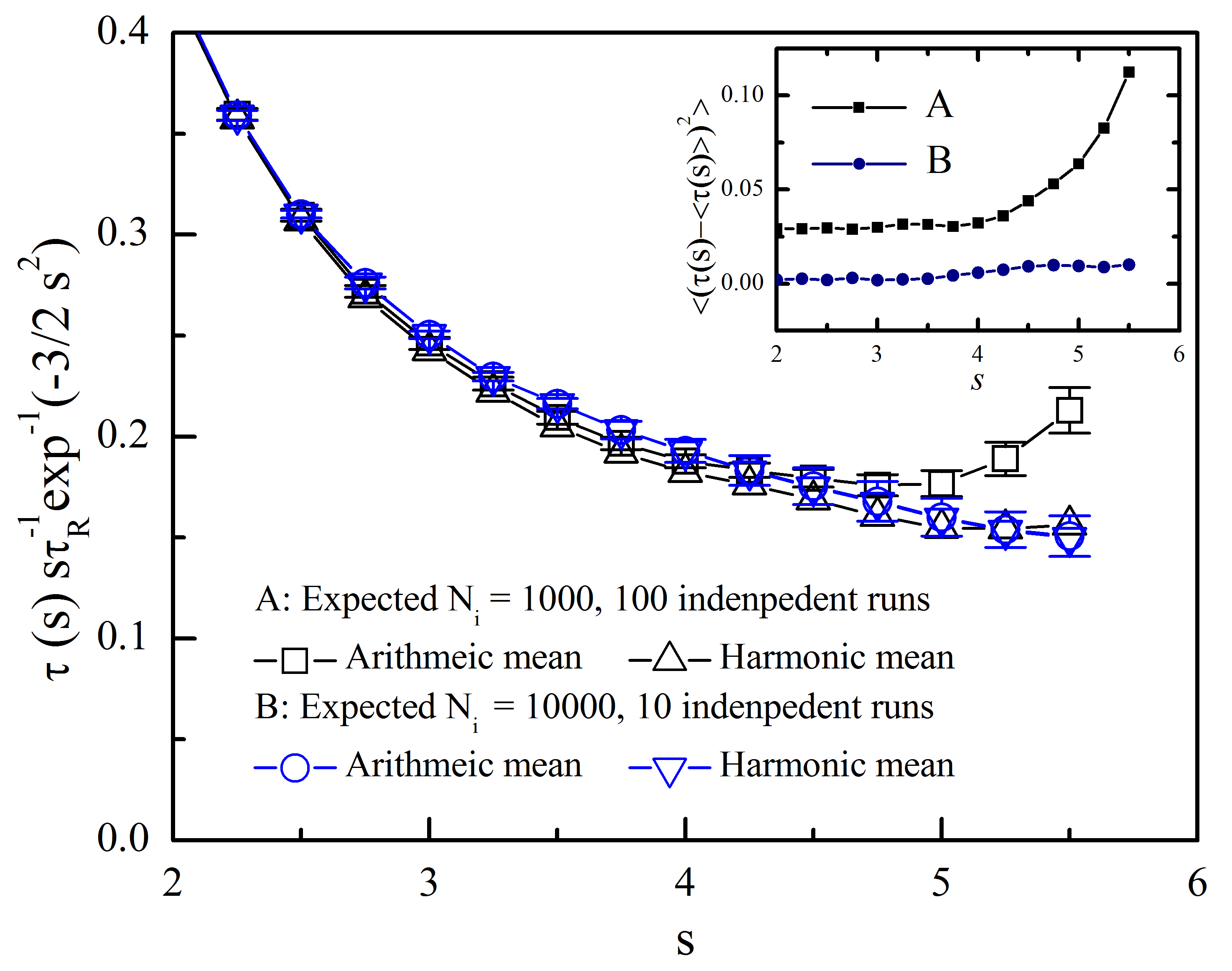}
\caption{A comparison between arithmetic mean and harmonic mean for averaging independent FFS runs. The simulations are carried out under two conditions, in condition A, expected $N_i$ is $1000$, and there are $100$ independent FFS runs, in condition B, expected $N_i$ is $10000$, and there are $10$ independent FFS runs. The inset is the variance for separate conditions.} \label{averageMethod}
\end{figure}

For discussion of the systematic and random error, we will consider the case of $N=32$ as an example. The independent simulations are carried on different CPUs, each CPU runs equal amount of FFS simulations. For all independent runs, the number of runs from each interface, $M_i$, must be identical, so that their statistical weights are equal. The time-step of $\Delta t= 0.01$ is chosen. Fig.\ref{averageMethod} compares the results from two sets of parameters. $M_i$ is estimated from an expected $N_i$, which are respectively $1000$ and $10,000$. For the first set, there are $N_{sim} = 100$ independent runs, while for the second, there are $10$ runs, such that the time-cost for both sets are equal. $M_i$ in first set is $10$ times smaller than in the second case, thus the variance is much bigger. By arithmetic mean, the average FP time is $\left<\tau(n)\right>= (1/N_{sim}) \sum_{i=0}^{N_{sim}} \tau(n)$, however, a systematic error can be detected after $z=4.5$, where $\tau$ suddenly goes up and diverges quickly from the average results of the other set, indicating that the direct average method is wrong for large $z$. In contrast, harmonic mean 
\begin{equation}
\left< \tau(n) \right> = \frac{N_{sim}}{\sum_{i=0}^{N_{sim}} 1/\tau_i(n)}.
\label{HarmonicMean}
\end{equation}
provides accurate results for all $z$.
In Figure \ref{averageMethod}, the FP time using this average method converge for both sets. This method allows us to simulate plenty of independent FFS runs, whose results can be averaged and are comparable to a simulation with larger $M_i$, while the time-cost in each processor would be much less.

\begin{acknowledgments}
This work was supported by the Engineering and Physical Sciences Research Council (EPSRC), grant EP/K017683.
\end{acknowledgments}

\bibliography{Jing_refs}

\end{document}